\documentclass[aps,preprint,groupedaddress]{revtex4-1}

\usepackage{graphicx}
\usepackage{amssymb}
\usepackage{graphicx}
\usepackage{helvet}
\usepackage{courier}
\usepackage{amsmath}
\usepackage{diagbox}
\usepackage{booktabs}
\usepackage{multirow}
\usepackage{subfig}
\usepackage{comment}
\usepackage{url}
\usepackage{array}
\usepackage{color}

\begin{document}


\title{\normalsize{Tales of Emotion and Stock in China: Volatility, Causality and Prediction}}


\author{Zhenkun Zhou$^1$, Ke Xu$^1$ and Jichang Zhao$^{2,\star}$}
\affiliation{$^1$State Key Lab of Software Development Environment, Beihang University\\
$^2$School of Economics and Management, Beihang University\\
$^\star$Corresponding author: jichang@buaa.edu.cn}


\date{\today}

\begin{abstract}
{\color{black}{How the online social media, like Twitter or its variant Weibo, interacts with the stock market and whether it can be a convincing proxy to predict the stock market have been debated for years, especially for China.}} As the traditional theory in behavioral finance states, the individual emotions can influence decision-makings of investors, it is reasonable to further explore these controversial topics systematically from the perspective of online emotions, which are richly carried by massive tweets in social media. Through thorough studies on over 10 million stock-relevant tweets and 3 million investors from Weibo, it is revealed that inexperienced investors with high emotional volatility are more sensible to the market fluctuations than the experienced or institutional ones, and their dominant occupation also indicates that the Chinese market might be more emotional as compared to its western counterparts. Then both correlation analysis and causality test demonstrate that five attributes of the stock market in China can be competently predicted by various online emotions, like disgust, joy, sadness and fear. Specifically, the presented prediction model significantly {\color{black}{outperforms the baseline model, including the one taking purely financial time series as input features}}, on predicting five attributes of the stock market under the $K$-means discretization. We also employ this prediction model in the scenario of realistic online application and its performance is further testified.

\end{abstract}

\keywords{Social Media, Stock Market, Sentiment Analysis, Volatility, Stock Prediction}

\maketitle

\section{Introduction}
\label{sec:introduction}

{\color{black}{There are a large number of tweets in popular platforms like Twitter and Weibo with explosive development of online social media.}} These tweets, spreading in terms of word-of-mouth, not only convey the {\color{black}{authentic}} information, but also reflect the emotional status of the authors. {\color{black}{For example, the development report from Weibo in 2015 officially shows the number of monthly active users is around 222 million.}} Around 100 million Chinese tweets are posted in Weibo every day and from which we cannot only sense what happens in China, but also how millions of users feel about their lives. The online social media indeed provide us an unprecedented opportunity to study the detailed human behavior from many new views.

{\color{black}{Investment decision and arbitrage model in the stock market attract much attention in recent decade.}} However, {\color{black}{there is much controversy as to whether}} online social media like Twitter can be excellent predictors, especially for the stock market in China~\cite{Bollen_predict_market,mao2014quantifying,wanyun2013investors}. {\color{black}{Different from western markets, the governmental intervention on the marketing regulations in China will introduce more non-market factors that might disturb the fluctuation of the stock market. Moreover,}} those possible interventions could be leaked through the social media and then greatly influence the investors' emotions and decisions. In the meantime, considering the irrationality of huge amount of individual {\color{black}{yet inexperienced}} investors in China (which is also rare in the west), their actions might be more easily affected by online news and other investors' feelings about the market. Then the messages about the stock market and the sentiments they convey could be {\color{black}{promising}} indicators for the market prediction. {\color{black}{Therefore}}, like the conventional behavioral finance theory claims, which the emotion can influence the decision-process of the investors, it is necessary to investigate the following important issues:
\begin{itemize}
\item How do different investors response emotionally to the performance of the stock market in China?
\item Is there indeed significant correlation between online emotions and attributes of the Chinese stock market?
\item Can online emotions predict the attributes of the stock market in China?
\item Which emotion does play the role in predicting various attributes of the Chinese stock market?
\end{itemize}

{\color{black}{In this study, we collect over ten million Chinese stock-relevant tweets from Weibo}} and classify them into five emotions, including anger, disgust, joy, sadness and fear. We discuss the emotional volatility of investors who belong to different categories classified by the number of followers and the gender. It's interesting to find that after a steep decline of the index, a short-term rising of market can help ordinary investors build the confidence quickly. However, the professional individual investors and institutional investors stay in fear and affirm it is the bear market. The volatility of online emotions of the experienced or institutional investors is far less than the ordinary inexperienced investors. In addition, female investors seem to possess higher emotional volatility than the male, suggesting their sensitiveness to the market fluctuation. Besides the daily closing index of Shanghai Stock Exchange (In the present paper, index refers in particular to Shanghai Stock Exchange Composite Index and the trading volume refers in particular to the daily volume of the Shanghai Stock Exchange), we consider the daily opening index, the intra-day highest index, the intra-day lowest index and the daily trading volume of the stock market. By both correlation analysis and Granger causality test, it is revealed that disgust has a Granger causal relation with the closing index, joy, fear and disgust have Granger causal relations with the opening index, joy, sadness and disgust have Granger causal relations with the intra-day highest and lowest index, and correlation between trading volume and sadness is unexpectedly strong. It's also surprising to find that anger in online social media possesses the weakest correlation or even is no relation with the Chinese stock market{\color{black}{, except for the experienced users (less than 2\% of the total investors yet less emotional as compared to others), for whom anger demonstrates a slightly significant connection of Granger causality with the market}}.

{\color{black}{According to these findings}}, we establish classification-oriented predictors, in which different emotions are selected as features, to predict five daily attributes of the stock market in China. The comparison with other baseline methods, {\color{black}{including the one taking purely financial time series as input features}} shows that our model can outperform them according to $K$-means discretization. And the model is also deployed in a realistic application and achieves the accuracy of 64.15\% for the intra-day highest index (3-categories) and the accuracy of 60.38\% for the trading volume (3-categories). Our explorations demonstrate that the online emotions, especially disgust, joy, sadness and fear, in Weibo indeed can predict the stock market in China.

\section{Related works}
\label{sec:related_works}

{\color{black}{Behavioral finance has provided solid proof that financial decisions are significantly driven by emotion and mood~\cite{dolan2002emotion,nofsinger2005social}. Nevertheless, because of the lacking}} of effective measurement method of emotions, stock prediction using emotions has been in dispute~\cite{NBERw13189,brown2004investor}. Traditionally, the general moods of investors are measured by surveys, e.g., Daily Investor Sentiment~\cite{Antweiler2004Is} and Investor Intelligence~\cite{asur2010predicting}. As psychological evidence {\color{black}{demonstrated}} that weather was associated with mood, weather was known as the proxy to examine the relation between emotion and stock market~\cite{Hirshleifer2003Good,Howarth1984A}. However, with the recent widespread presence of personal computers and Internet, public emotions can be extracted easily from data on online platforms. Using Twitter as a corpus, researchers built sentiment classifiers, which are able to determine different sentiments for tweets~\cite{baseline_method,pak2010twitter,twitter_various_techniques}. Especially on Sina Weibo platform, Zhao et al. trained a fast Naive Bayes classifier for Chinese emotion classification, which is now available online for temporal and spatial sentiment pattern discovery~\cite{moodlens}. {\color{black}{These works indeed provide foundations for considering public emotions as input for stock predictions.}}

{\color{black}{There have long been discussed on the predictive power of social media aiming at different fields~\cite{gayo2012wanted,sakaki2010earthquake}}}. In the field of finance, Bollen et al. found that public mood on Twitter can predict the Dow Jones Industrial Average\cite{Bollen_predict_market}. Meanwhile, the public mood dimensions of Calm and Happiness seemed to have a predictive effect. However, the tweets they collected were associated with whole social status, not just the stock market in America, which could not {\color{black}{well}} represent online investors' sentiment. Oh et al. also showed stock micro-blog sentiments did have predictive power for market returns~\cite{oh2011investigating}. Instead of emotion on social media, some researchers examined textual representations in financial news articles for stock prediction~\cite{Schumaker2009bu,cohen2013mood}. Antweiler and Frank found the messages posted on Yahoo! Finance and Raging Bull help predict market volatility, however the effect on stock returns is economically small~\cite{Antweiler2004Is}. Ding et al. proposed a deep learning method for event-driven stock market prediction on large-scale financial news dataset~\cite{XiaoDing:2015uo,leinweber2011event}. Besides, Bordino et al. showed that daily trading volumes of stocks traded in NASDAQ100 were correlated with daily volumes of queries related to the same stocks~\cite{bordino2012web}, as well as the volumes of queries at the weekly level~\cite{preis2010complex}.

{\color{black}{In comparison to this extensive literature, much less research focused on the relationships between the Chinese stock market and online emotions in social media.}} Mao et al. pointed out that Twitter did not have a predictive effect {\color{black}{with respect to}} predicting developments in Chinese stock markets~\cite{mao2014quantifying}. They advised adopting the tweets on Weibo platform to research Chinese stock market. Based on 66,317 tweets of Weibo with one year and two emotion categories, Cheng and Lin found that the investors' bullish sentiment of social media can help to predict trading volume of the stock market, but still {\color{black}{did}} not work for the market returns~\cite{wanyun2013investors}. Because of less collected data set and simple emotion classification, it is not easy to generalize their conclusions to other realistic scenarios.

{\color{black}{In light of these studies}}, we focus purely on the stock market in China and try to understand the volatility, causality and predictive ability of multiple online emotions in Weibo. Different from the previous studies, we {\color{black}{attempt}} to develop predictors from more data sets and more sentiments and to predict more attributes of the real market.

{\color{black}{This paper is an extended version of \cite{Zhou2016Can}. The supplementary content to \cite{Zhou2016Can} is stated as follows. From the perspective of online emotions, we investigate how investors respond online to the stock market. Investors are divided into multiple categories according to the number of followers and gender in Weibo. We define the ratio of joy to fear ($RJF$) to reflect the overall psychological state of different investors while trading in greed and fear. The deviation of $RJF$ can be further used to model the emotional volatility of investors. It is the first to reveal that inexperienced investors are more sensible to the market fluctuations than the experienced or institutional ones from the perspective of social media, suggesting the Chinese market is emotional. In realistic application, we evaluate the performance of models based on online emotions extracted from different categories of users. We also build the baseline models based on market return to examine the robustness of online emotions' predictive power.}}

\section{Data sets}
\label{sec:data_sets}

In this section, {\color{black}{the data sets will be depicted. We obtain over 10 million public stock-relevant tweets on Weibo through its open APIs}} over one year from December 1st 2014 to December 7th 2015. Employing the Bayesian classifier of emotions and other preprocessing methods, we obtain time series which could reflect the online emotion of daily stock market in the period. The stock market data referring to the index and trading volumes is crawled and processing in the same period.

\subsection{Online stock market emotions}
\label{online_emotions}
{\color{black}{The Emotions of investors can be obtained}} through many different approaches, like questionnaires in previous study. With the explosive development of the social media in China, more and more investors express {\color{black}{their feelings}} on Weibo. Therefore, we choose and utilize the characteristic of Weibo to collect the online emotion referring to Chinese stock market.

From December 1st 2014 to December 7th 2015, the massive public tweets on Weibo are collected through its open APIs. However, only a fraction of the tweets {\color{black}{is}} semantically related with Chinese stock market. Filtering out the irrelevant tweets and remaining the data that truly represents the stock market emotion is a very significant step. Therefore, we manually select six Chinese keywords, including Stock, Stock Market, Security, The Shenzhen Composite Index, The Shanghai Composite Index and Component Index with help of expertise from the background of finance. These manually selected keywords are supposed to depict the overall status of Chinese stock market sufficiently. We postulate that if the text of tweet contains one or more of the six selected keywords, the tweet describes the news, opinions or sentiments about Chinese stock market. In our database, the number of tweets related to stock market, involving one or more keywords, is a total of 10,550,525 from December 1st 2014 to December 7th 2015.
 
\begin{figure}
\centering
\includegraphics[height=6cm]{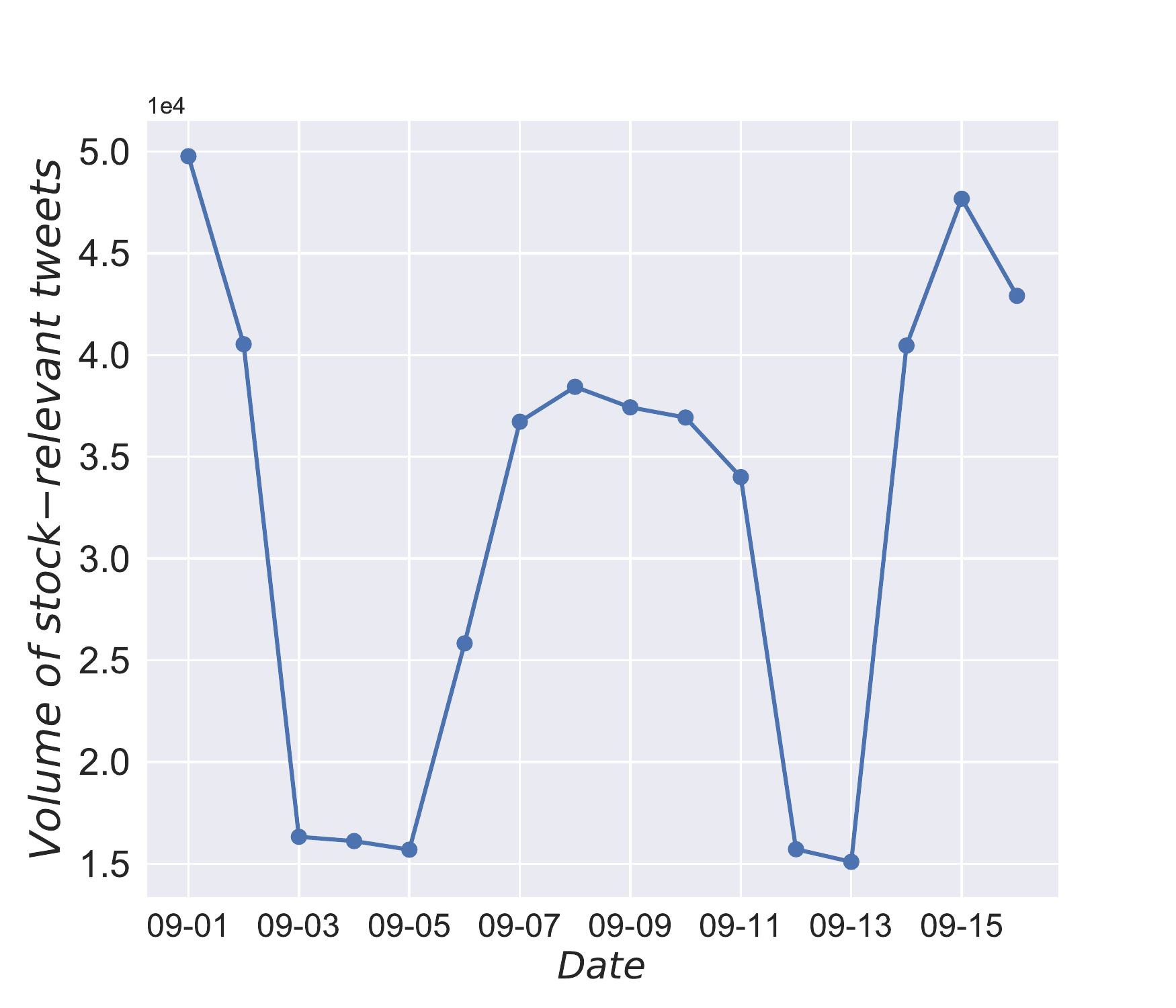}
\caption{{\color{black}{Volume of the stock-relevant tweets from September 1st to 16th in 2015. There are respectively Memorial Day between 9-3 and 9-5 and a weekend between 9-12 and 9-13, which are also non-trading days.}}}
\label{fig:amount_of_tweets}
\end{figure}

{\color{black}{It is widely accepted that there are six basic emotions for human, including joy, surprise, fear, sadness, anger and disgust~\cite{ekman2013emotion}. Hence in our previous research of emotion classification~\cite{moodlens}, emoticons are mapped manually by several experts into these six sentiments, serving as the class labels of tweets. However, the emotion `surprise' is absent from Weibo and there is no sufficient convincing emoticon corresponding to it. Therefore, we choose the other five types of online emotions to represent the emotional expression in Weibo and investigate their connections with the stock market of China. In total we collect over 3.5 million emotionally labeled tweets as the corpus and train a fast Naive Bayes classifier, with an empirical precision of 64.3\% to automatically categorize a tweet without emoticon into a certain emotion.}} We arrange the stock-relevant tweets with one day as the time unit and employ the sentiment classifier to label them with the emotions. There are five online emotion time series: $X_{anger}, X_{disgust}, X_{joy}, X_{sadness}$ and $X_{fear}$. Online emotions are represented by $X = (X_{anger}, X_{disgust}, X_{joy}, X_{sadness}, X_{fear}).$

{\color{black}{We find the volume of tweets reduces significantly on non-trading days.}} Fig.~\ref{fig:amount_of_tweets} shows the volume of the tweets related to the stock market from September 1st to 16th in 2015. There are separately Memorial Day between September 3rd and 5th and a weekend between September 12th and 13th, which are both non-trading days. We consider that online stock market emotion on non-trading days could not help us analyze and predict Chinese stock market. Hence, removing the data items on non-trading days from the time series, the results retain significant emotion data. It also partly reflects that tweets selected by the keywords could {\color{black}{well represent fluctuations of}} the stock market.

\begin{figure}
\centering
\includegraphics[width=8cm]{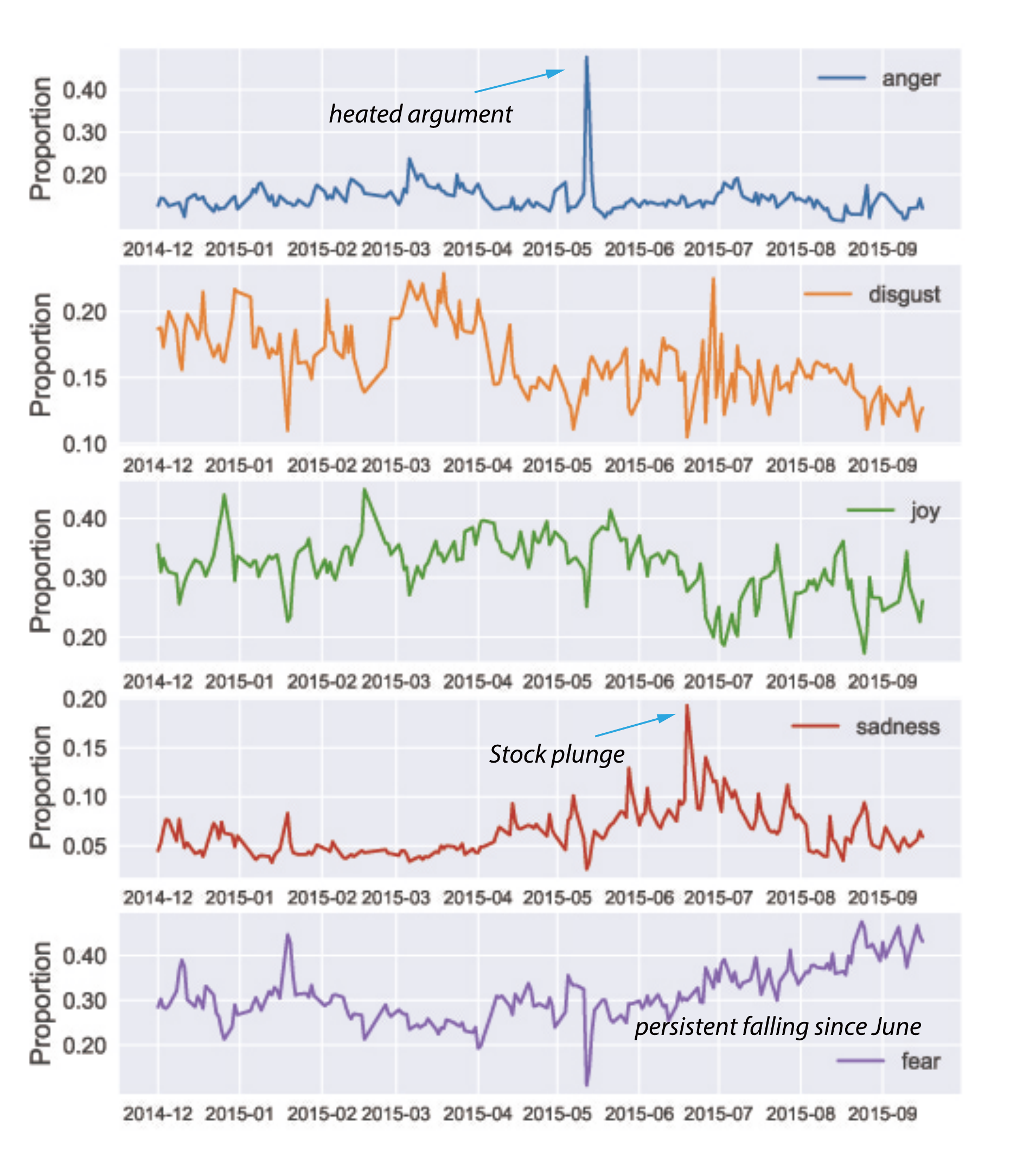}
\caption{{\color{black}{Time series of each online stock market emotion from December 1st 2014 to September 16th 2015.}}}
\label{fig:emotion}
\end{figure}

For the sake of stability of online emotion data, we measure the relative value (proportion) of each mood on one day as the final online stock market emotion $X$. Fig.~\ref{fig:emotion} shows online stock market emotion time series $X$ from September 1st to 16th in 2015. We observe the spike in $X_{anger}$ on May 12th in 2015, when there is a heated argument between CEOs of listed companies. On June 19th, 2015, there was a plunge in Chinese stock market, with a fall of the index with 6.41\% and $X_{sadness}$ arrived the maximum. Since June 2015, persistent falling of the index caused inward fears of investors, which can be seen from the sharp growth of $X_{fear}$ in Fig.~\ref{fig:emotion}.

From the above observations, it can be concluded that the fluctuation of the sentiments can be connected with remarkable events in the stock market. It further inspires us to investigate the correlation and even causality between emotions and the market, which will provide the foundation for the predicting models.

\subsection{Stock market data}

\begin{figure}
\centering
\includegraphics[width=10cm]{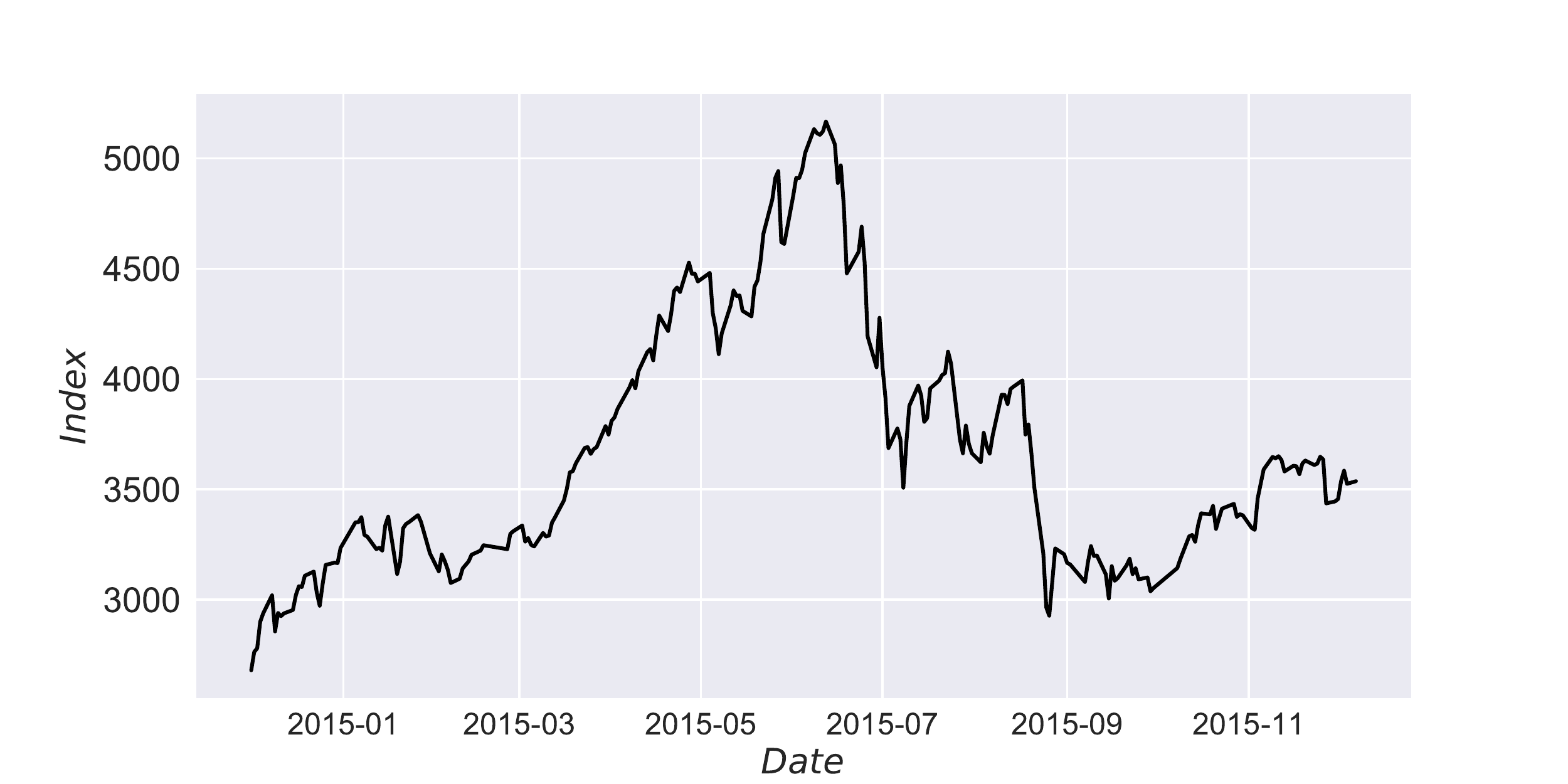}
\caption{{\color{black}{Shanghai Stock Exchange Composite Index from December 2014 to December 2015.}}}
\label{fig:sh}
\end{figure}

{\color{black}{The economists and traders consider the Shanghai Stock Exchange Composite Index as reflecting the overall status of the Chinese stock market.}} Therefore, the index (shown in Fig.~\ref{fig:sh}) is selected as price attribute of the stock market to analyze and predict. In particular, there are four values in candlestick charts of the index, which are respectively the closing index, the opening index, the intra-day highest index, the intra-day lowest index. We transform the values of the index into $Close$, $Open$, $High$ and $Low$ (to express rate of change on $i$-th day), and they can be written as
\begin{gather}
\begin{split}
 Close_{i} = \frac{Index_{close,i}-Index_{close,i-1}}{Index_{close,i-1}} \times 100, \\
 Open_{i} = \frac{Index_{open,i}-Index_{close,i-1}}{Index_{close,i-1}} \times 100, \\
 High_{i} = \frac{Index_{high,i}-Index_{close,i-1}}{Index_{close,i-1}} \times 100, \\
 Low_{i} = \frac{Index_{low,i}-Index_{close,i-1}}{Index_{close,i-1}} \times 100.
\end{split}
\end{gather}

In addition to these four attributes, the trading volume of Shanghai Stock Exchange is also a key target used to reflect the status of the Chinese stock market. The time series of trading volume on each day is not transformed at all. 

We crawl historical data of the index and trading volume from December 1st 2014 to December 7th 2015. In this period, the number of trading days is totally 249 in our research. As a result, we obtain five time series which depict stock market's state on each day including $Y_{close}, Y_{open}, Y_{high}, Y_{low}$ and $Y_{volume}$. Each time series is a column vector of $Y$ (shown in Fig.~\ref{fig:discretiztion}), i.e., $Y = (Y_{close}, Y_{open}, Y_{high}, Y_{low}, Y_{volume}).$

The dataset ($X$ and $Y$) is divided into two parts according to the date: the 80\% data for training (from December 1st 2014 to September 16th 2015) and the 20\% data for testing (from September 17th to December 7th in 2015). The training set is used to not only analyze the relation between online emotions and the stock market but also fit and estimate the prediction model. The testing set is kept in a vault and brought out only at the end of evaluation in realistic application.

\section{Volatility of online emotions from different investors}
\label{sec:volatility}

To explore the connection between emotion and stock, the first intuition is to understand {\color{black}{how investors respond}} online to the market from the perspective of emotions. Hence in this section, we divide the investors into multiple categories according to the number of followers and gender in Weibo. {\color{black}{Note that the followers and gender information of the users can be extracted from the pubic profiles returned by Weibo's open  APIs.}} Then the ratio of joy to fear ($RJF$) is defined to reflect the averaged psychological state of different investors while trading in greed and fear and its deviation can be further used to model the emotional volatility of investors. The comparison of emotional volatility of {\color{black}{different investors demonstrate}} that professional or institutional investors are more emotionally stable than ordinary investors as the stock index rises or declines. More interestingly, female investors might be more sensible to the market performance as compared to the male.

\subsection{Investor categorizations}

Massive individual or institutional investors express their reflections to the market performance through emotional tweets in Weibo. Though it is difficult to determine the investment level or type of Weibo users from their online profiles and stock-related tweets, the number of followers in Weibo is indeed a convincing indicator to measure users' influence to their online community~\cite{icwsm10cha,zhang2012motivations,networkeffect}. Generally{\color{black}{,}} users with higher numbers of followers will be more likely to be influenced by altruistic messages and calls to help others and offers to join discussion groups and opinion forums. Referring to investment and exchange in the stock market, it appears that the investors who join Weibo to actively participate in discussions with inspiring traders and provide help and information to the community, can offer potential values to other participants and thus enjoy higher popularity in Weibo. For example, the stock market experts or successful investors are always followed by plenty of other investors in Weibo because of their expertise and experience in market analysis and sometimes they even recommend certain strategies on the targeted stocks to followers. In fact{\color{black}{,}} Weibo also provides a feature of bonus that followers who eventually profit from online suggestions or recommendations can voluntarily reward those experts. This feature greatly encourages the stock information sharing online and the influential investors with more experience will gain more attention, like followers in Weibo. Note that here the investors mentioned is filtered out from the Weibo users if posting stock-related tweets and we assume that investors of real-world can be evenly sampled through this way{\color{black}{, especially considering the vast number of Weibo users}}.

\begin{figure}
\centering
\includegraphics[width=8cm]{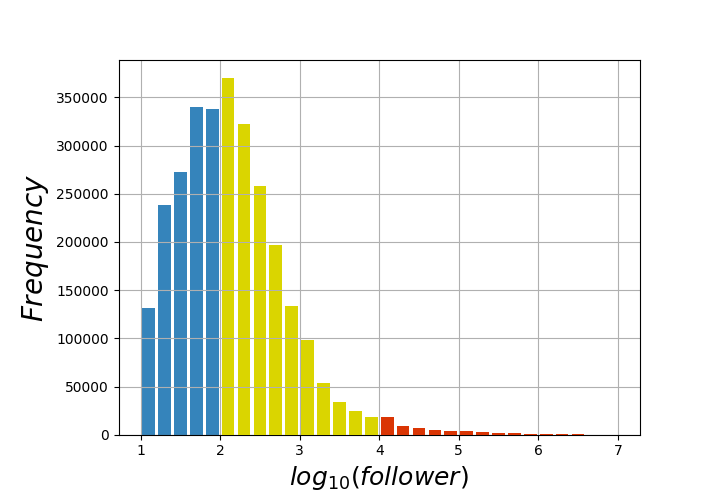}
\caption{Distribution of the number of followers (denoted as $n_f$) across all Weibo users. We highlight in blue the bottom 53.5\% of the area, corresponding to users with low number of followers ($n_f<=10^2$) in F-level I category. Red bars indicate users with high number of followers (top 1.5\%, $n_f>=10^4$) in F-level III category while yellow corresponds to the remaining 45\% users with medium number of followers ($10^2<=n_f<=10^4$) in F-level II category.}
\label{fig:dist_follower}
\end{figure}

Inspired by the above observations, in this study we divide the investors into three categories according to their numbers of followers in Weibo, including F-level I, II and III as shown in Fig.~\ref{fig:dist_follower} (1.8 million F-level I users, 1.5 million F-level II users and 5.5 thousand F-level III users). Specifically, the number of followers in F-level I users is less than $10^2$ (around the Dunbar's number~\cite{dunbarnumber} and possible for any user) and they occupy the bottom 53.5\% of the investors we filter out in Weibo. The number of followers in F-level II users ranges between $10^2$ and $10^4$ (a large volume and infeasible for most of users~\cite{guo2011sina}) and they occupy the medium 45\%. {\color{black}{We consider that most of F-level I and F-level II users are inexperienced ordinary investors.}} The left users with more than $10^4$ followers belong to the F-level III and they only occupy 1.5\%. In fact{\color{black}{,}} most of users in F-level III are either sophisticated and professional individual investors in stock investment or the online agents of institutional investors.

It is found that emotion expression is demographics dependent~\cite{association_theory,hu2016ambivalence} and thus gender matters in investigating the emotional volatility of investors. In the meantime, because of the common fact that decision-makings can be affected by demographics like gender, we further divide investors in Weibo into categories of female (1.4 million) and male (2 million) based on user profiles, offering more perspectives in understanding investors' online emotion expression.

\begin{figure}
\centering
\subfloat[Anger]{\includegraphics[width=5.5cm]{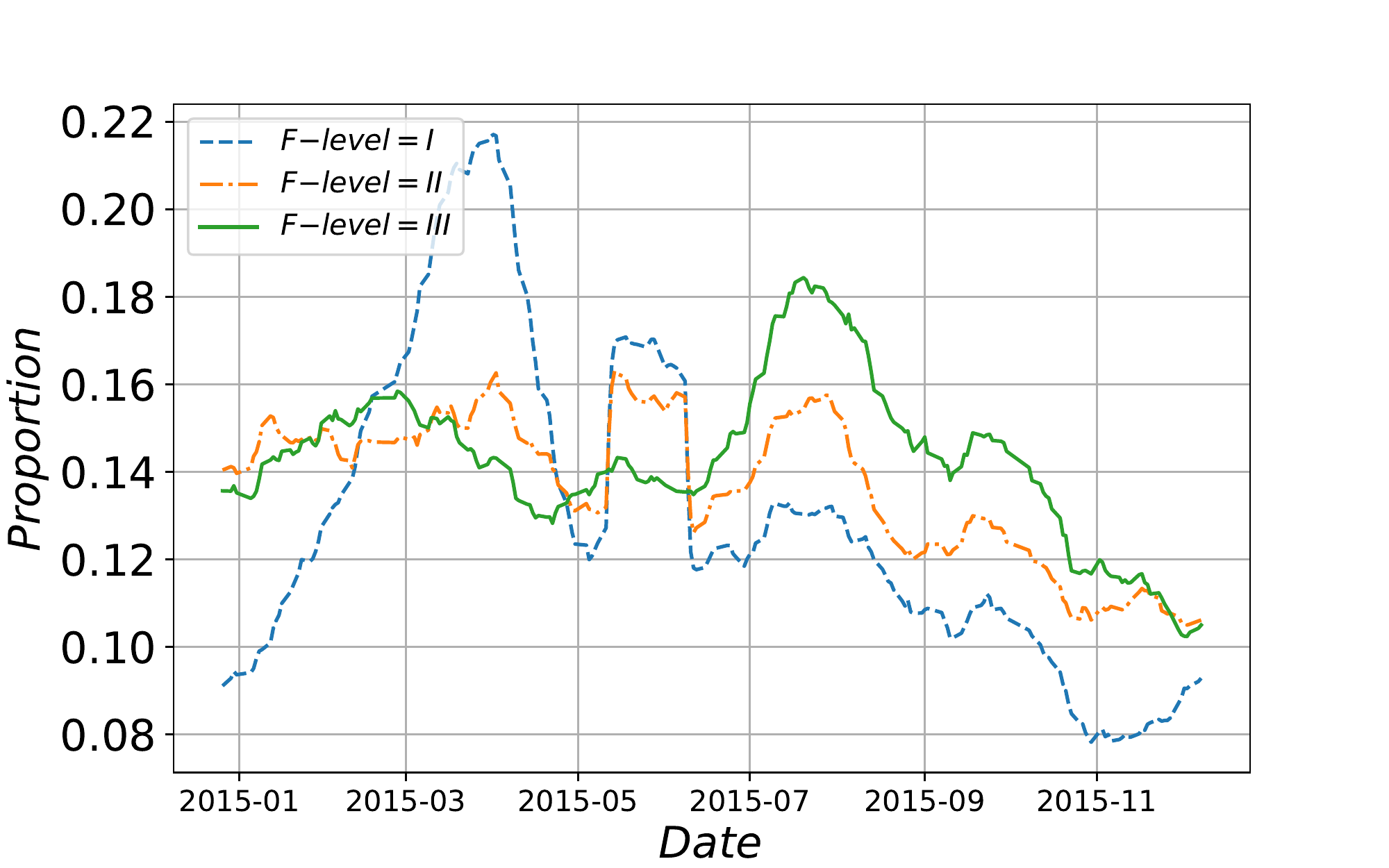}
\label{fig:emotion-0}}
\hfil
\subfloat[Disgust]{\includegraphics[width=5.5cm]{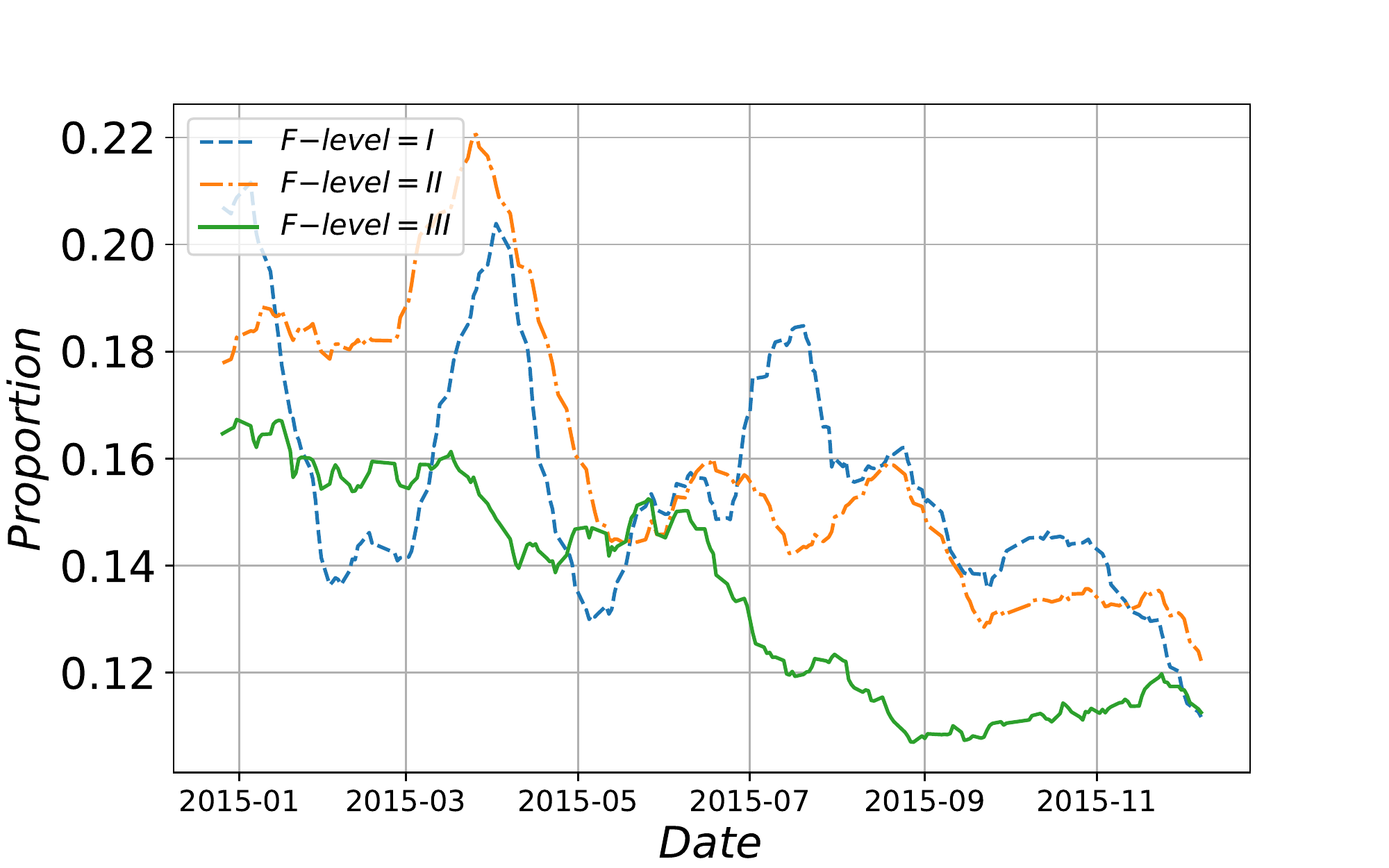}
\label{fig:emotion-1}}
\hfil
\subfloat[Joy]{\includegraphics[width=5.5cm]{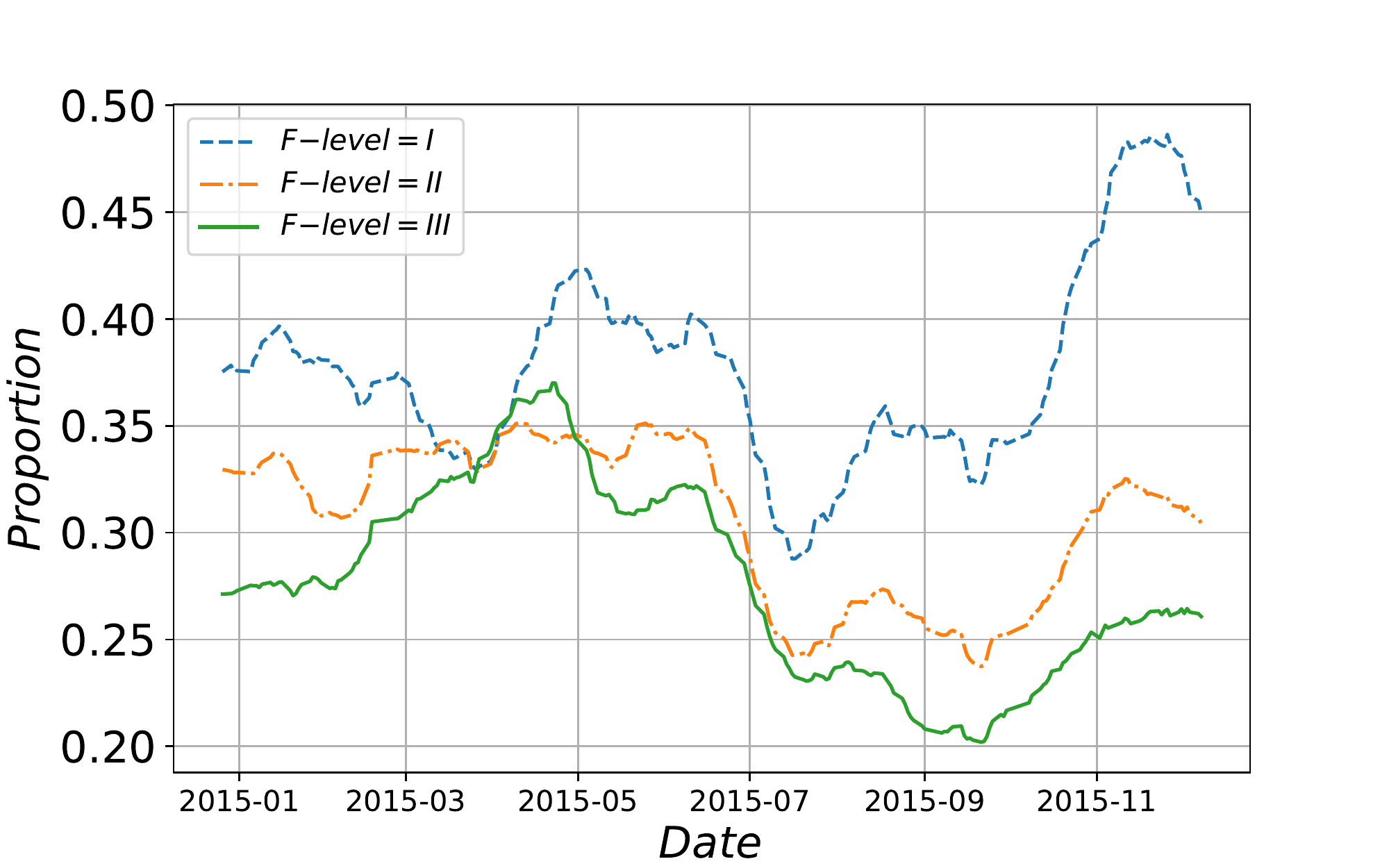}
\label{fig:emotion-2}}
\hfil
\subfloat[Sadness]{\includegraphics[width=5.5cm]{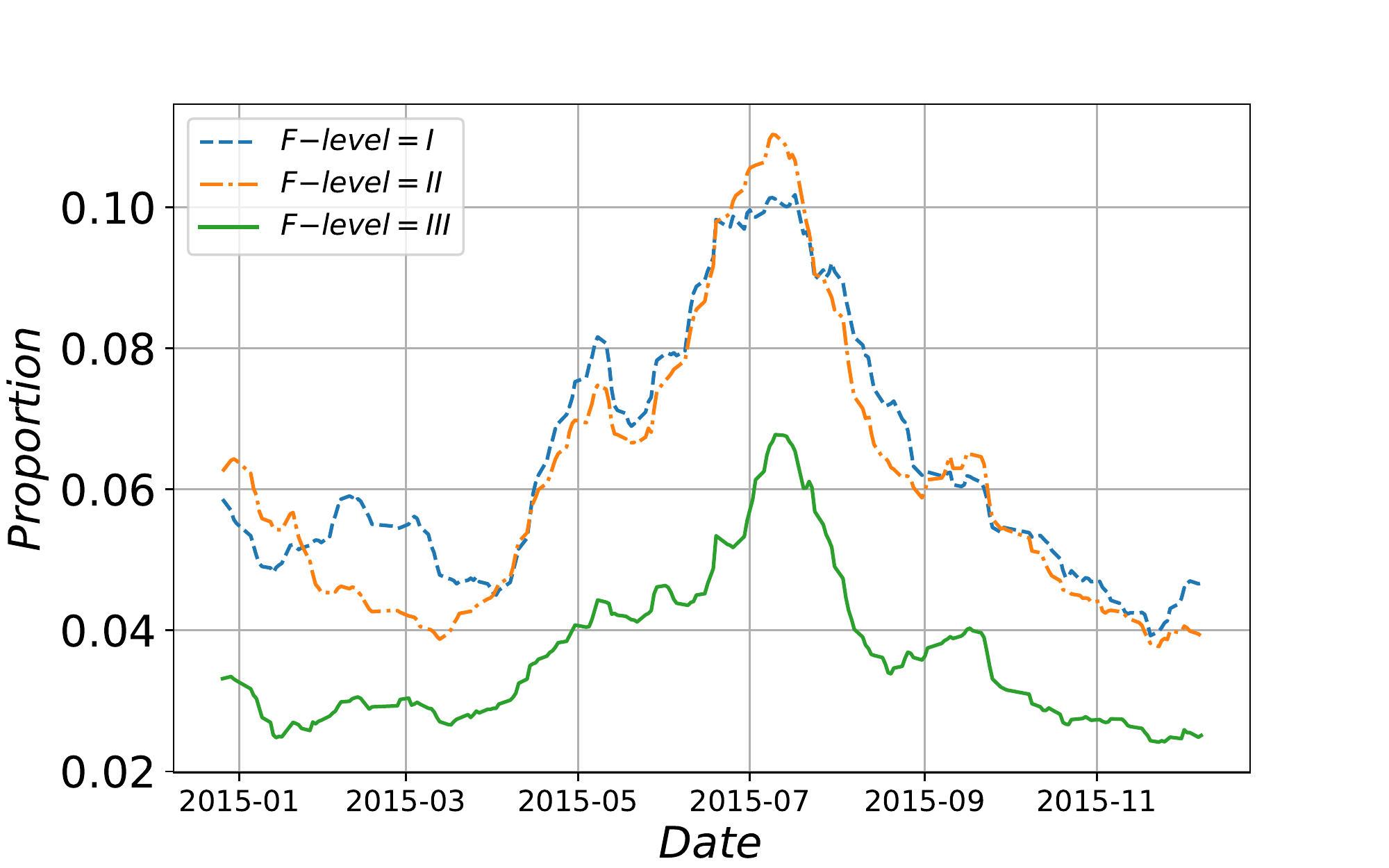}
\label{fig:emotion-3}}
\hfil
\subfloat[Fear]{\includegraphics[width=5.5cm]{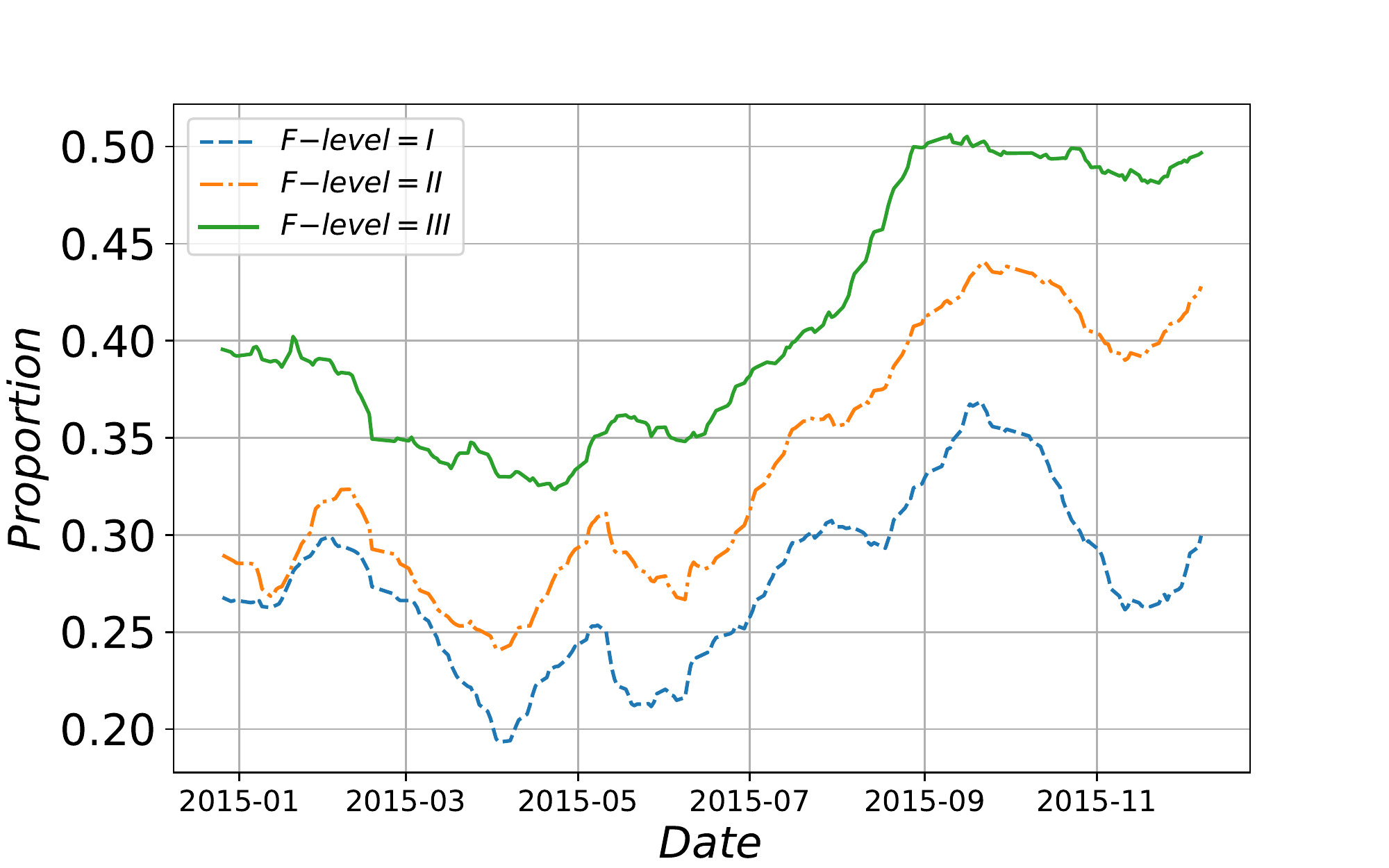}
\label{fig:emotion-4}}
\caption{{\color{black}{Five online emotions rolling with a window of 20 days from December 1st, 2014 to December 7th, 2015 for different categories of investors.}}}
\label{fig:emotion_F-level}
\end{figure}

In section~\ref{online_emotions}, we define the online stock market emotions $X$ without considering the category of users. Adopting the same method, we obtain three or two groups of $X$ based on F-level or gender categories. Focusing on the centralized tendency of $X$ in categories, we calculate the moving average value of $X$, in which the sliding window size is fixed to 20 (days). Fig.~\ref{fig:emotion_F-level} exhibits the time series (moving average $X$) rolling with a window of 20 days of each online stock market emotion for F-level categories. We find that F-level I investors' emotions vary more obviously than that of other two categories, especially for $X_{joy}$ and $X_{fear}$. Surprisingly, to the F-level III users that composed mainly by experienced or institutional investors, $X_{fear}$ takes the dominant proportion after May 28th, 2015, indicating their affirmation of bearish after the steep decline of Chinese market.

\subsection{Trading in greed and fear}
Greed and fear are two intrinsic psychological states relating to fluctuation of stock market. And thus in general, the trading activity of the agents can be psychologically characterized by greed and fear. In specific, with respect to greed, the investors optimistically believe in rising markets and thus buy stocks; while regarding to fear, if stock prices fluctuate too abruptly, they panic and sell stocks. In addition, the agents switch between these two psychological states and thus demonstrate a pattern of trading in greed and fear. It has also been revealed that as historical volatility of the market is low, investors are rather calm and vice versa~\cite{Westerhoff2004Greed}. One of the best available and accepted tools to measure stock market volatility is CBOE Volatility Index {\color{black}{(VIX)}}, elaborated by Chicago Board Options Exchange in 1993. In other words, VIX can be defined as a sentiment ratio of Wall Street's greed to fear~\cite{nossman2009vix}. It is usually used by traders to check the degree of investor complacency or market fear. However, we argue that online emotions can be used directly and accurately to measure the sentiment ratio of greed to fear, instead of the volatility of market indirectly and vaguely. Because in the circumstance of bullish market, investors intend to post positive tweets with joyful sentiment and thus joy can be an ideal proxy of the external emotional expression of greed and complacency for investors. Then based on online emotions of investors, we define the ratio of joy (greed) to fear ($RJF$) at $t$ moment as 
\begin{equation}
 RJF_{t} = \frac{X_{joy,t}}{X_{fear,t}},
\end{equation}
which is a new and purely emotion-based index to describe the psychological state of investors in Chinese stock market. While $RJF_{t}$ is greater than 1, the investors consider the market is rising and are pleased to put more money into it. However, investors are irrational and become greedy as the $RJF$ is too high. While $RJF_{t} < 1$, fear is the dominant emotion in the market and investors are afraid of the loss of benefit. In addition, the higher proportion of investors are bears who sell stocks in expectation of a drop of price. The investors stay rational and the emotions are stable as $RJF$ is around 1.

\begin{figure}
\centering
\subfloat[F-level I]{
\includegraphics[width=5.5cm]{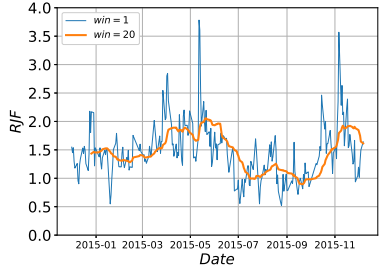}
\label{fig:RJF_F-level_I}
}
\hfil
\subfloat[F-level II]{
\includegraphics[width=5.5cm]{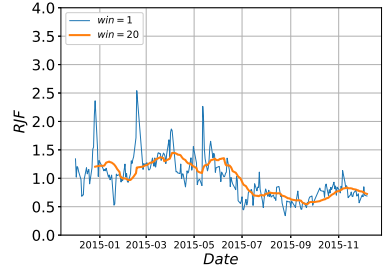}
\label{fig:RJF_F-level_II}
}
\hfil
\subfloat[F-level III]{
\includegraphics[width=5.5cm]{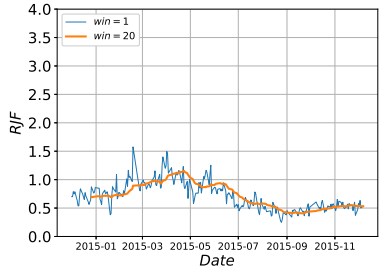}
\label{fig:RJF_F-level_III}
}
\hfil
\subfloat[Female]{
\includegraphics[width=5.5cm]{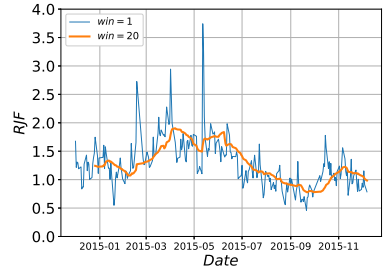}
\label{fig:RJF_female}
}
\hfil
\subfloat[Male]{
\includegraphics[width=5.5cm]{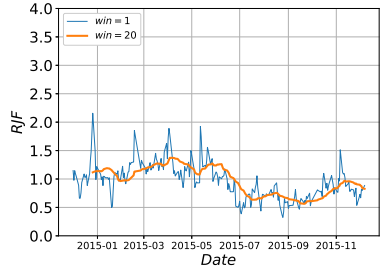}
\label{fig:RJF_male}
}
\caption{$RJF$ of investors in F-level and gender categories. The blue lines show the daily $RJF$. The orange lines {\color{black}{show}} the moving average of corresponding $RJF$ curve with a 20 days window.}
\label{fig:RJF}
\end{figure}

How $RJF$ varies with time for investors of F-level and gender categories is shown in Fig.~\ref{fig:RJF}, in which the original $RJF$ is portrayed by the blue curves and its moving average $RJF$ with a fixed window of $20$ days is plotted in orange. According to $RJF$'s variation, we find interesting emotional characteristics of different investors. As can be seen in Fig.~\ref{fig:RJF_F-level_I}, $RJF$ of F-level I gradually increases until middle of May 2015 and arrives at the peak of 3.78. However, it then falls rapidly because a steep decline of the index. And with the recovery of Chinese stock market, F-level I's $RJF$ rises again during October and November 2015. The $RJF$ of F-level II demonstrates a similar evolving trend with F-level I before August, but investors in F-level II do not recover their confidence with $RJF$ being less than $1.0$ almost throughout after October. Different from investors in both F-level I and II, investors in F-level III shows a continuous trend of $RJF<1$, especially after the middle of May 2015 and then only fluctuates slightly around 0.5, suggesting that experienced individual investors and institutional investors lastingly affirm the market is bearish after the steep decline in June 2015. 

\subsection{Emotional volatility of investors in multiple categories}
The difference in $RJF$ of diverse investors can be further probed though the emotional volatility, which is used to represent the stability of the confidence to the market. And lower emotional volatility accordingly suggests higher emotional stability. In previous study, volatility (referring in particular to historical volatility in this paper) is a statistical measure of the dispersion of returns for a given stock or market index and it can be simply measured by using the standard deviation~\cite{Schwert1989Why}. Using the same idea with stock volatility, here we define the emotional volatility $\sigma_{R}$ (call $RJF$ $R$ for short) as
\begin{gather}
\centering
\begin{split}
 \mu_{R} = \frac{1}{N}\sum_{t=1}^{N}R_{t}, \\
 \sigma_{R} = \sqrt{\frac{\sum_{t=1}^{N}(R_{t} - \mu_{R})^2}{N-1}}.
\end{split}
\end{gather}

\begin{figure}
\centering
\subfloat[]{\includegraphics[width=4cm]{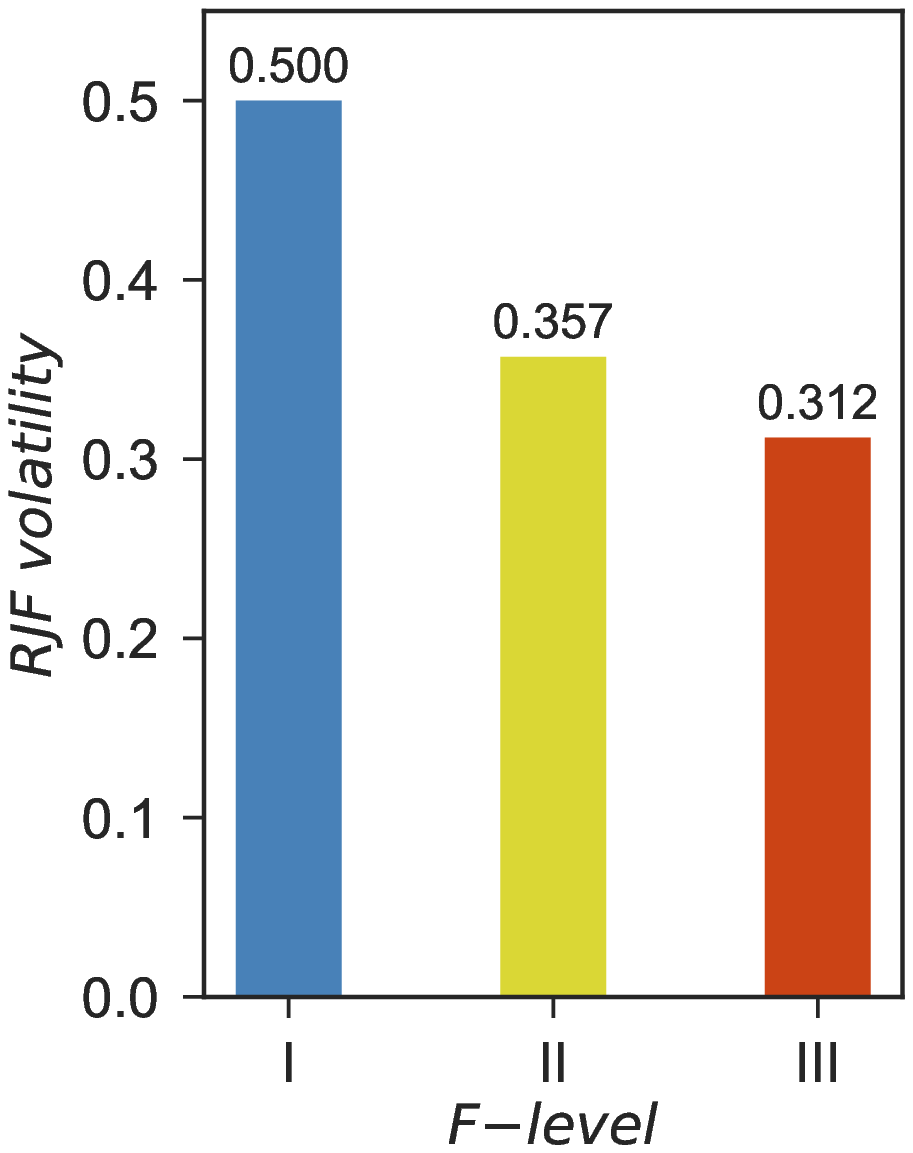}
\label{fig:vol_F-level}}
\hfil
\subfloat[]{\includegraphics[width=4cm]{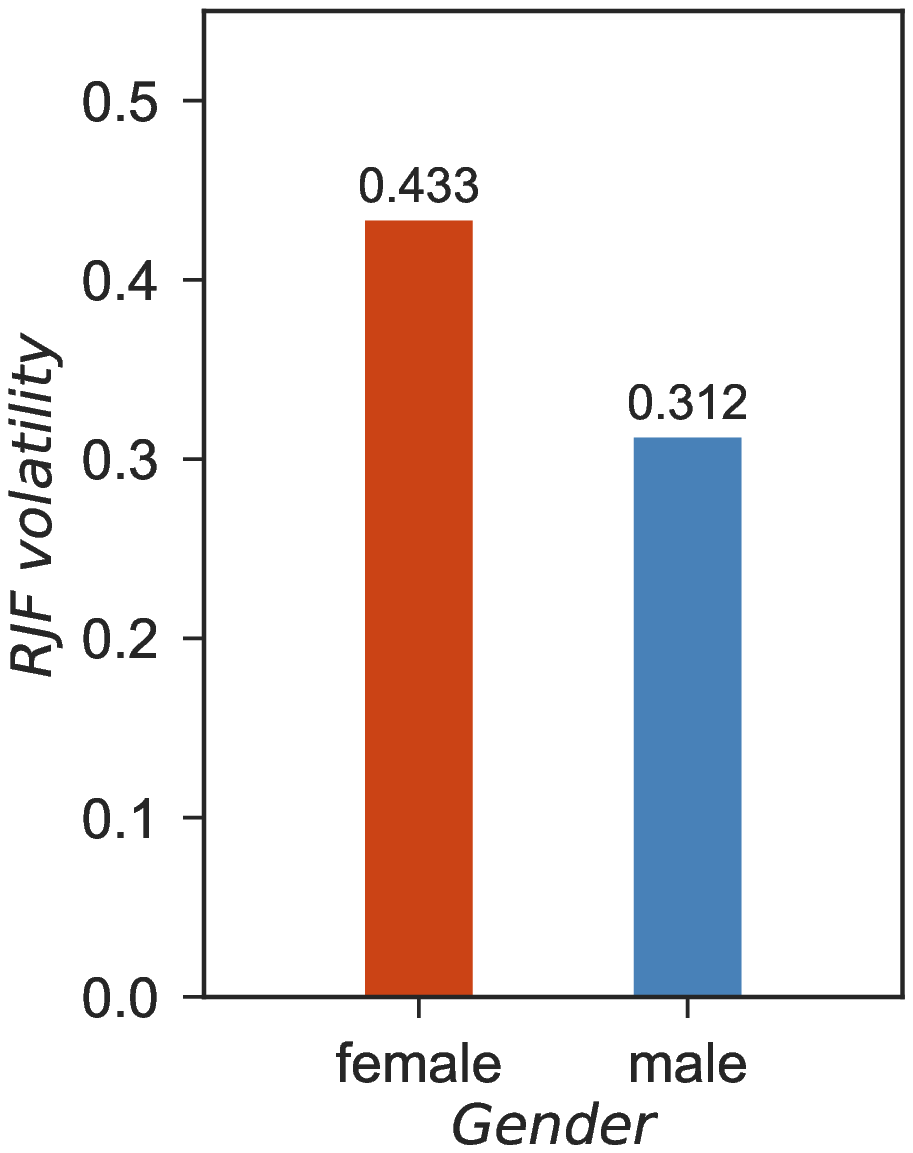}
\label{fig:vol_gender}}
\hfil
\subfloat[]{\includegraphics[width=7cm]{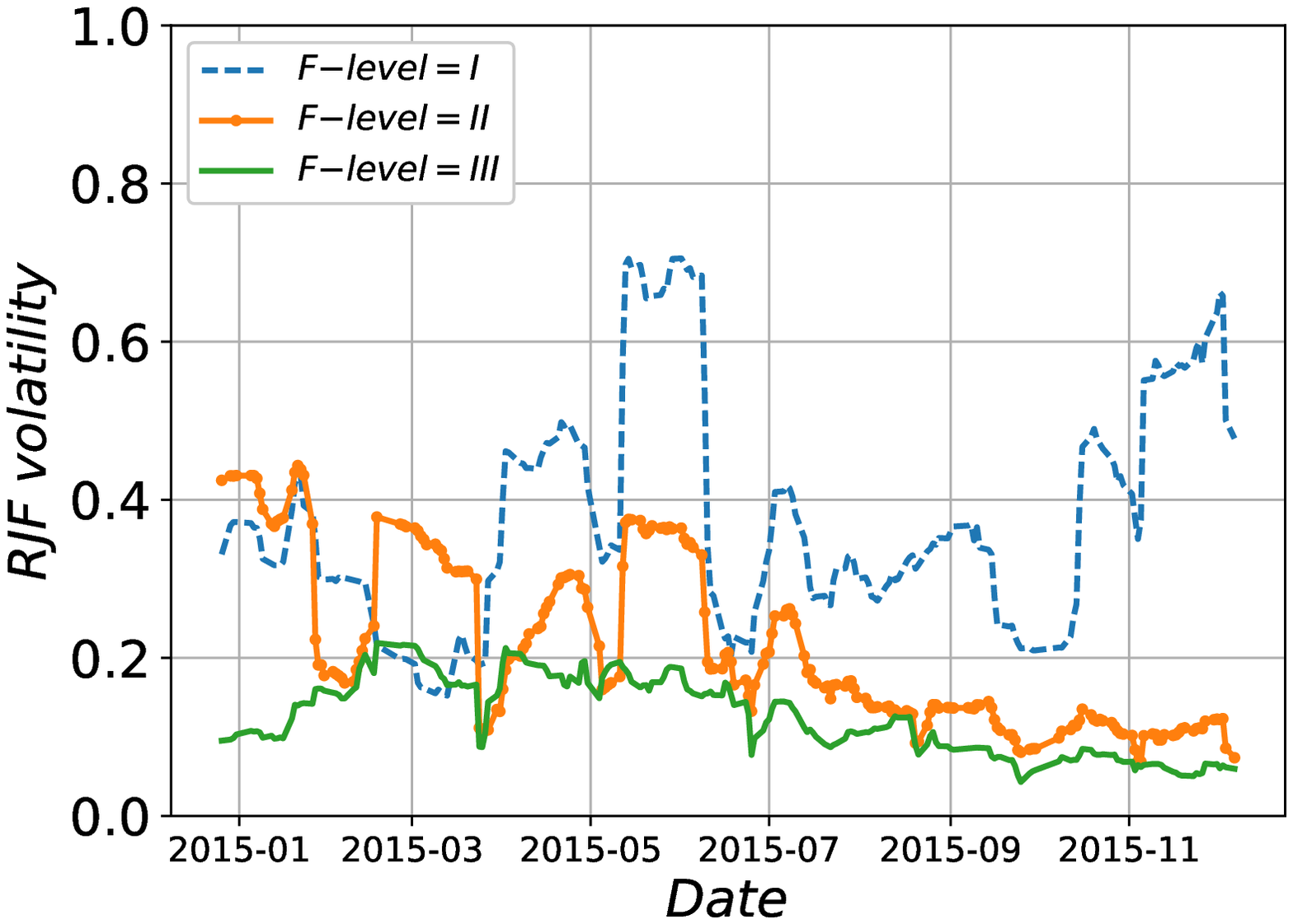}
\label{fig:vol_F-level_win}}
\hfil
\subfloat[]{\includegraphics[width=7cm]{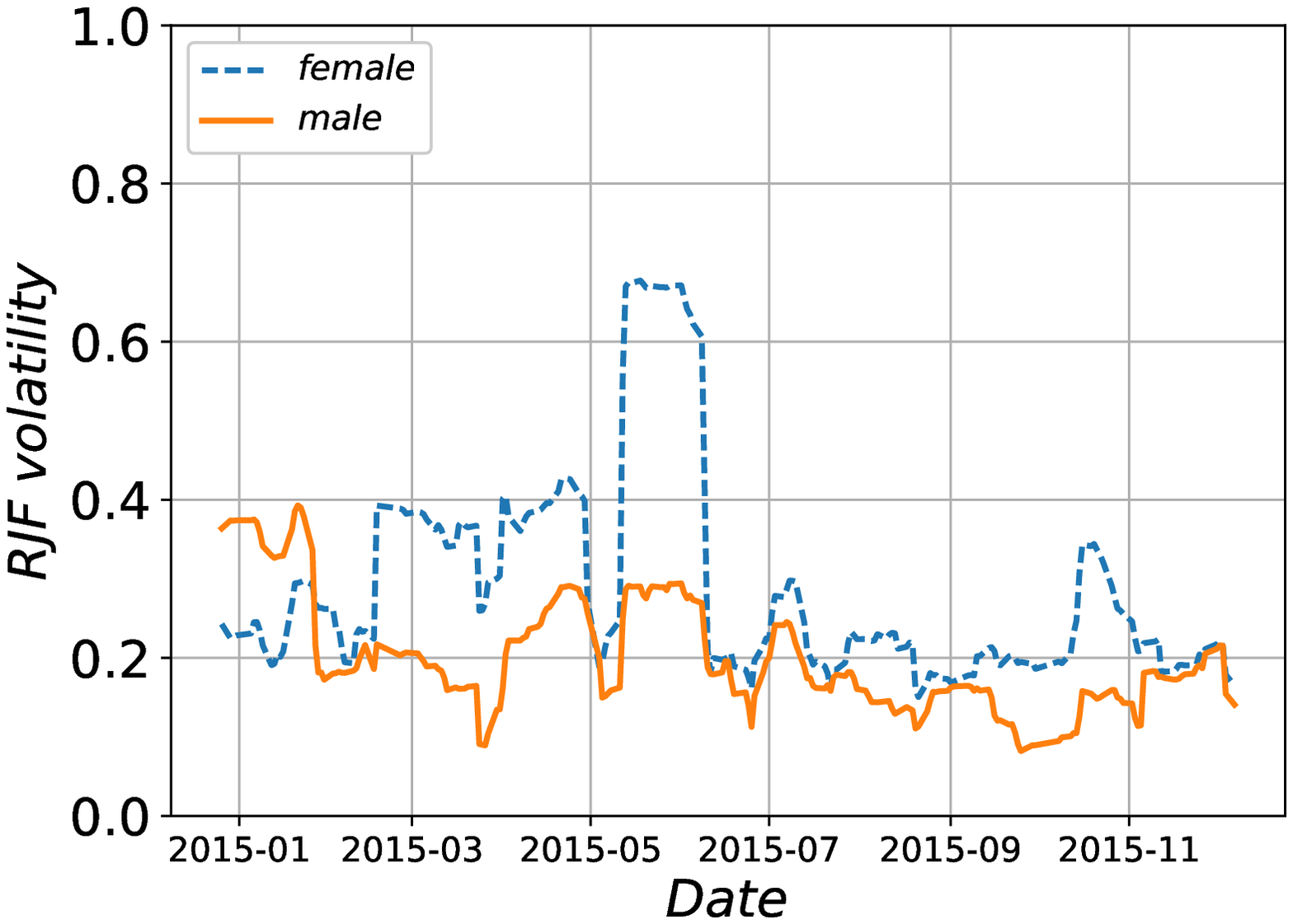}
\label{fig:vol_gender_win}}
\caption{(a, b) Global emotional volatility and (c, d) the moving average of emotional volatility with a window of 20 days.}
\label{fig:vol}
\end{figure}

Considering the data set in this study covering a stage of dramatic change of the stock market in China (see Fig.~\ref{fig:sh}), the emotional volatility will indeed offer more insights in profiling investors' emotional response to different market performance. We measure $\sigma_{R}$ for investors of different F-levels and gender, and the global $RJF$ volatilities are shown in Fig.~\ref{fig:vol}. The global $RJF$ volatilities of F-level I, II and III are respectively $0.500$, $0.357$ and $0.312$. {\color{black}{Accordingly}}, it can be concluded that the newbies of F-level I are sensitive to the fluctuation of stock market, however, experienced investors belonging to F-level III are more emotionally stable. Conventional wisdom leads us to believe that women are more emotional than men~\cite{parkins2012gender,deng2016gender} and $RJF$ of female and male consistently indicate the same insight that women are more emotionally expressive in the realm of Weibo referring to the stock, possessing a higher global $RJF$ volatility of $0.433$ (which is only $0.312$ for the male).

To highlight the variation of $RJF$ volatility with stock market swing, we calculate the moving average $RJF$ volatility with the window of 20 days which are shown in Figs.~\ref{fig:vol_F-level_win} and \ref{fig:vol_gender_win}. Referring to the F-level I investors, the $RJF$ volatility is greater than 0.65 during peak periods from May 13th to June 8th, 2015. In the meanwhile, the index increase to 5000 which it has never reached in the past eight years. Afterwards, the index begins falling rapidly to near 3300 on July 2015. The $RJF$ volatility recovers to 0.65 on December 3rd, and then the index falling again to near 3000 in January 2016. The moving average $RJF$ volatility of F-level II users is below 0.45 and the peak value is above 0.4 between December 2014 and January 2015. However, the volatility continues falling with the turbulence of stock market. Referring to F-level III investors, they are less emotional than other investors and the maximum is even less than 0.22. They make it clear that the bear is coming before the plunge in Chinese stock market. While from the perspective of gender, online emotions of women fluctuate widely between May and June in 2015, and the emotional volatility is over 0.6; for male investors, the volatility is almost below 0.3 and demonstrates a more stable response.

To sum up, by dividing investors into different groups, their emotional response is systematically investigated from the perspective of $RJF$ and emotional volatility. It is surprisingly disclosed that inexperienced investors, i.e., users with small numbers of followers in Weibo, are more sensible to the market fluctuation than the experienced or institutional ones. Considering the fact that inexperienced investors taking a dominant partition (over 98\% in our data set) of the Chinese market, their high volatility in confidence to the market might greatly magnify the market's signals of growths or declines and results in collective but emotional buys or sells, which will function negatively on the market itself. That's to say, the Chinese market might be significantly emotional. These findings encourage us to further investigate the correlation and even causality between emotion and stock in the following section. 

\section{Correlation and causality between emotions and stock market}
\label{sec:cor_cau}

It is intuitive that the index of stock market will influence investors' emotions, for example, rising index brings joy but declining index produces fear or sadness. However, whether emotions of investors function on the fluctuation of the market, positively or negatively, still remains unclear. Considering the dominant occupation of inexperienced users in the market, we argue that the stock market in China is significantly emotional and the connection between emotions and the market deserves further explorations. In this section, by thoroughly investigating the statistical correlations between emotions and market attributes, we demonstrate that emotions like joy and fear possess causal relations with the stock market and can be promising features in developing prediction models for stocks in China. {\color{black}{It is worth noting that considering the tiny fraction of experienced or institutional investors (less emotional) in the market, we do not discuss the correlation or causality for different investors in the following parts of the study.}}

\subsection{Correlation between emotions and stock market}
\label{sec:correlation}

In the above section, we define the two groups of time series: $X$ (represents online stock market emotions) and $Y$ (represents the stock market), which contribute to discuss the correlation between online emotions and the stock market. However, the purpose of the paper is to find out whether online emotions can predict the stock market in China. {\color{black}{In finance, the short-term stock prediction according to the past five trading days have the highest explanatory power~\cite{krauss2017deep}. Supposing that online emotions in past five trading days are available for short-term stock prediction, we shift emotion series to an earlier date: 1 to 5 days.}} Hence, each emotion corresponds to 5 time series according to shifted time. Each category of online emotions can be defined as (the categories of emotions are represented by $e$, $e=anger, sadness, joy, disgust$, or $fear$) $X_{e} = (X_{e, 1}, X_{e, 2}, X_{e, 3}, X_{e, 4}, X_{e, 5}).$

For the analysis the relation of $X$ and $Y$ ($T$ represents one certain time series of $X$ or $Y$), we normalize all the time series, in which data items are transformed to the values from 0 to 1 as $T_{i}=(T_{i} - T_{min})/(T_{max} - T_{min})$. $T_{i}$ is the $i$-th item in time series $T$, $T_{max}$ is the maximal value of $T$, and $T_{min}$ is the minimal value of $T$. Then, by using Pearson correlation analysis, we measure the linear dependence between $x$ ($X_{e, t}$, the emotion $e$ ahead of $t$ days in $X$) and $y$ (one target of $Y$) as $\rho$ is the Pearson correlation coefficient of time series $x$ and $y$ which is defined as $\rho = \frac{\Sigma(x_{i} - \bar{x})(y_{i} - \bar{y})}{\sqrt{\Sigma(x_i - \bar{x})^2\Sigma(y_i - \bar{y})^2}}.$

\begin{figure}
\centering
\subfloat[$Y_{close}$ and $X$]{\includegraphics[width=0.28\textwidth]{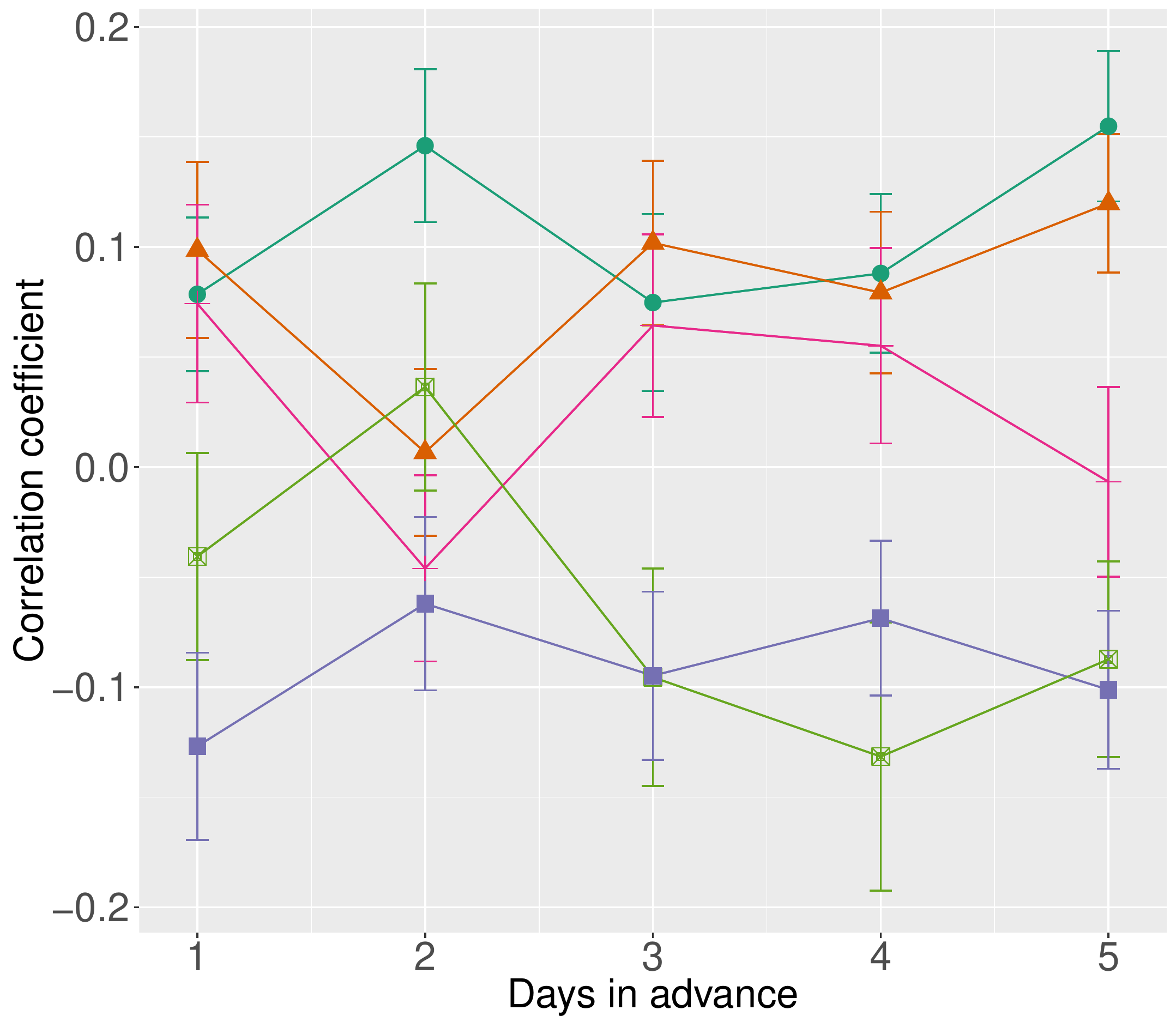}}
\hfil
\subfloat[$Y_{open}$ and $X$]{\includegraphics[width=0.28\textwidth]{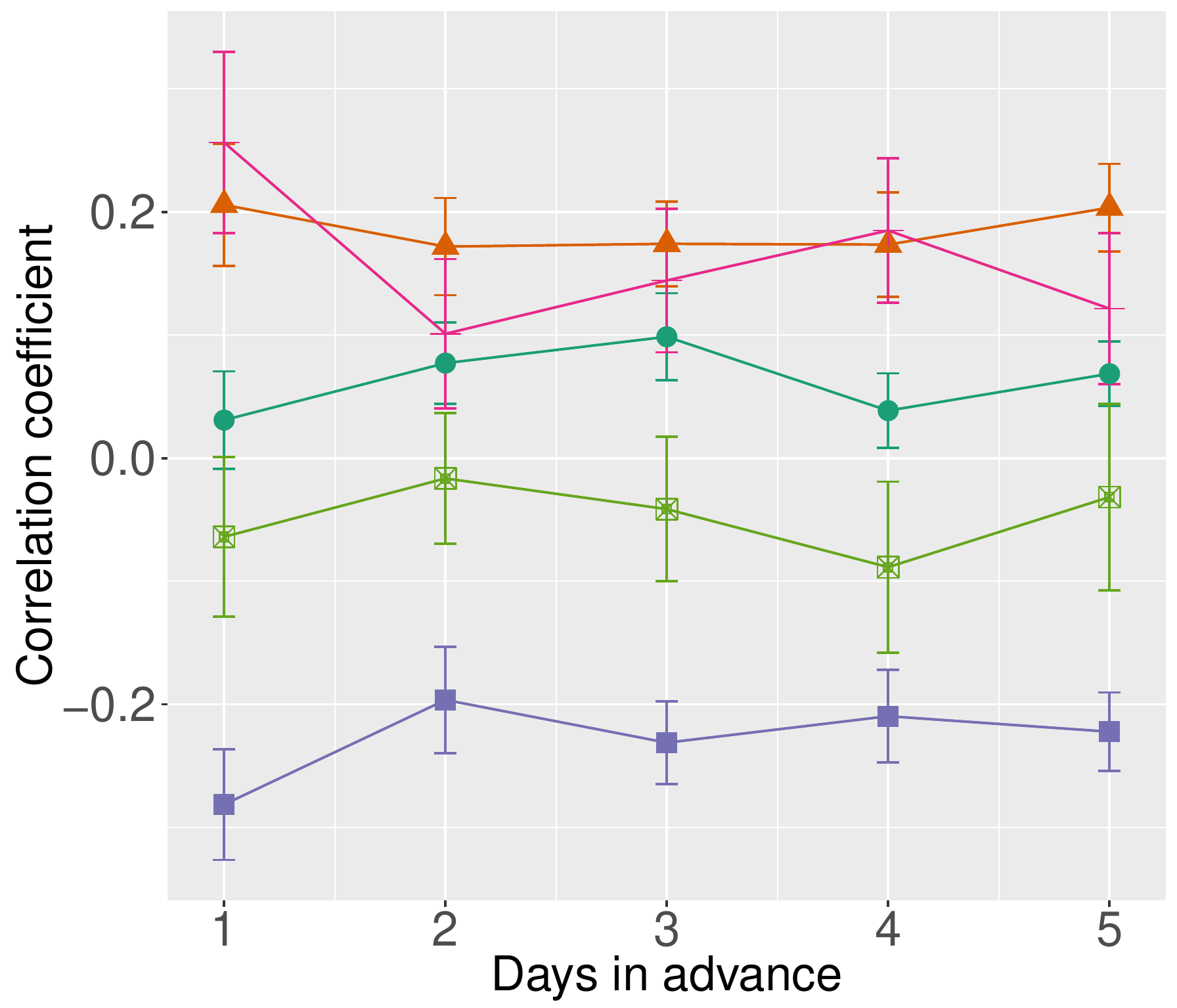}}
\hfil
\subfloat[$Y_{high}$ and $X$]{\includegraphics[width=0.28\textwidth]{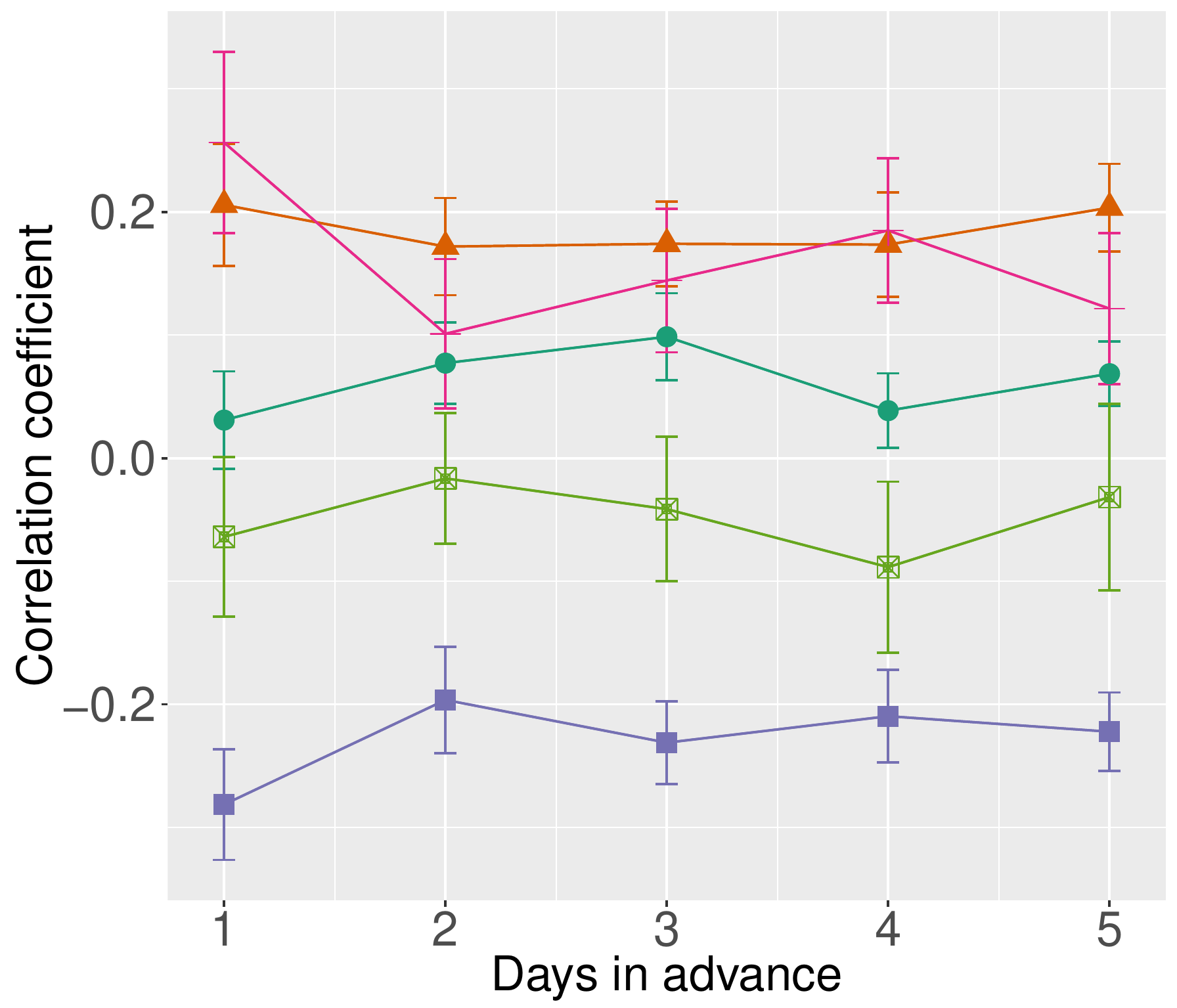}}
\hfil
\subfloat[$Y_{low}$ and $X$]{\includegraphics[width=0.28\textwidth]{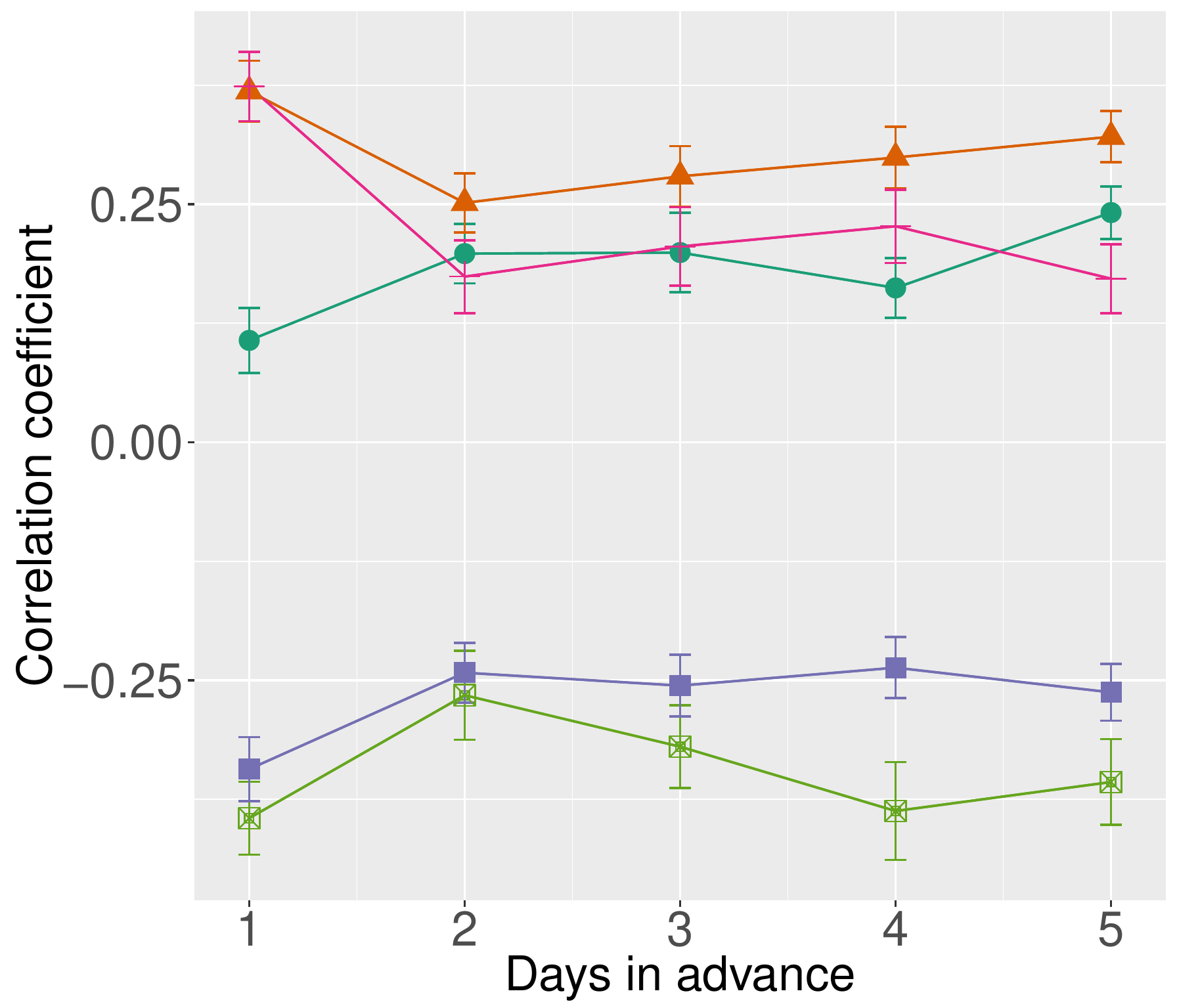}}
\hfil
\subfloat[$Y_{volume}$ and $X$]{\includegraphics[width=0.33\textwidth]{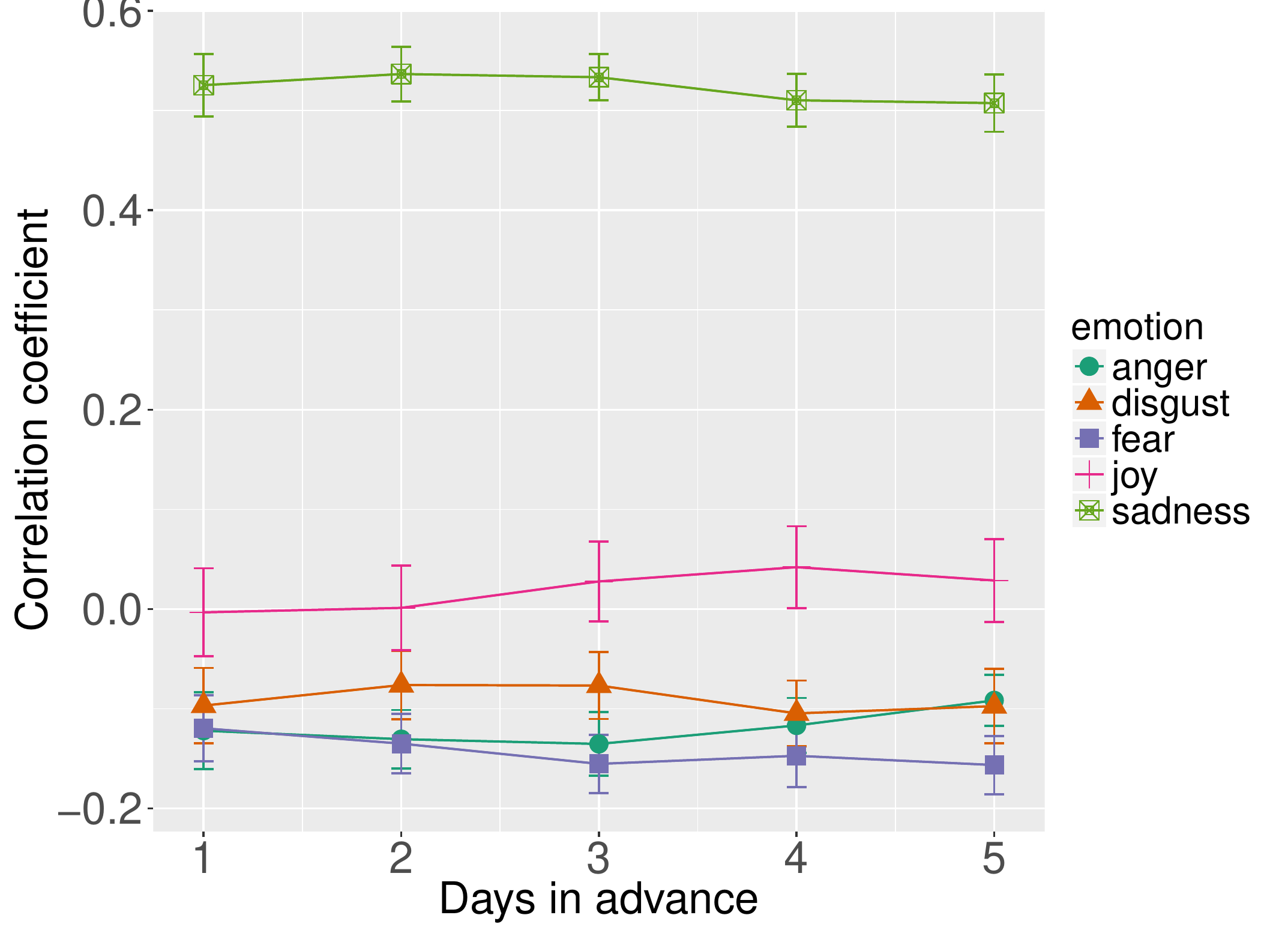}}
\caption{{\color{black}{Pearson correlation coefficient between five targets of stock market and online emotion time series.}}}
\label{fig:error_bar}
\end{figure}

\begin{figure}
\centering
\subfloat[$Y_{close}$ and $X$]{\includegraphics[width=0.28\textwidth]{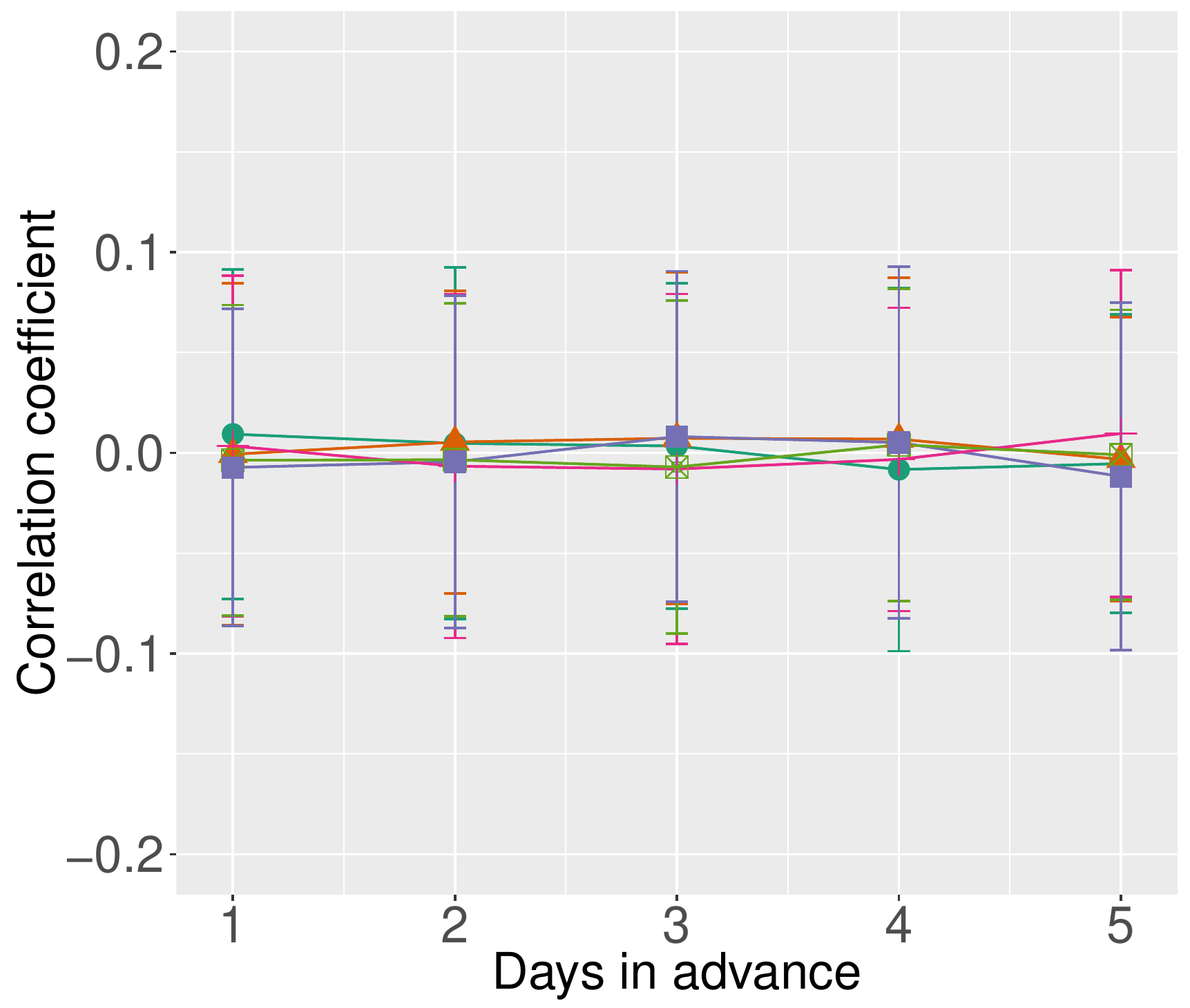}}
\hfil
\subfloat[$Y_{open}$ and $X$]{\includegraphics[width=0.28\textwidth]{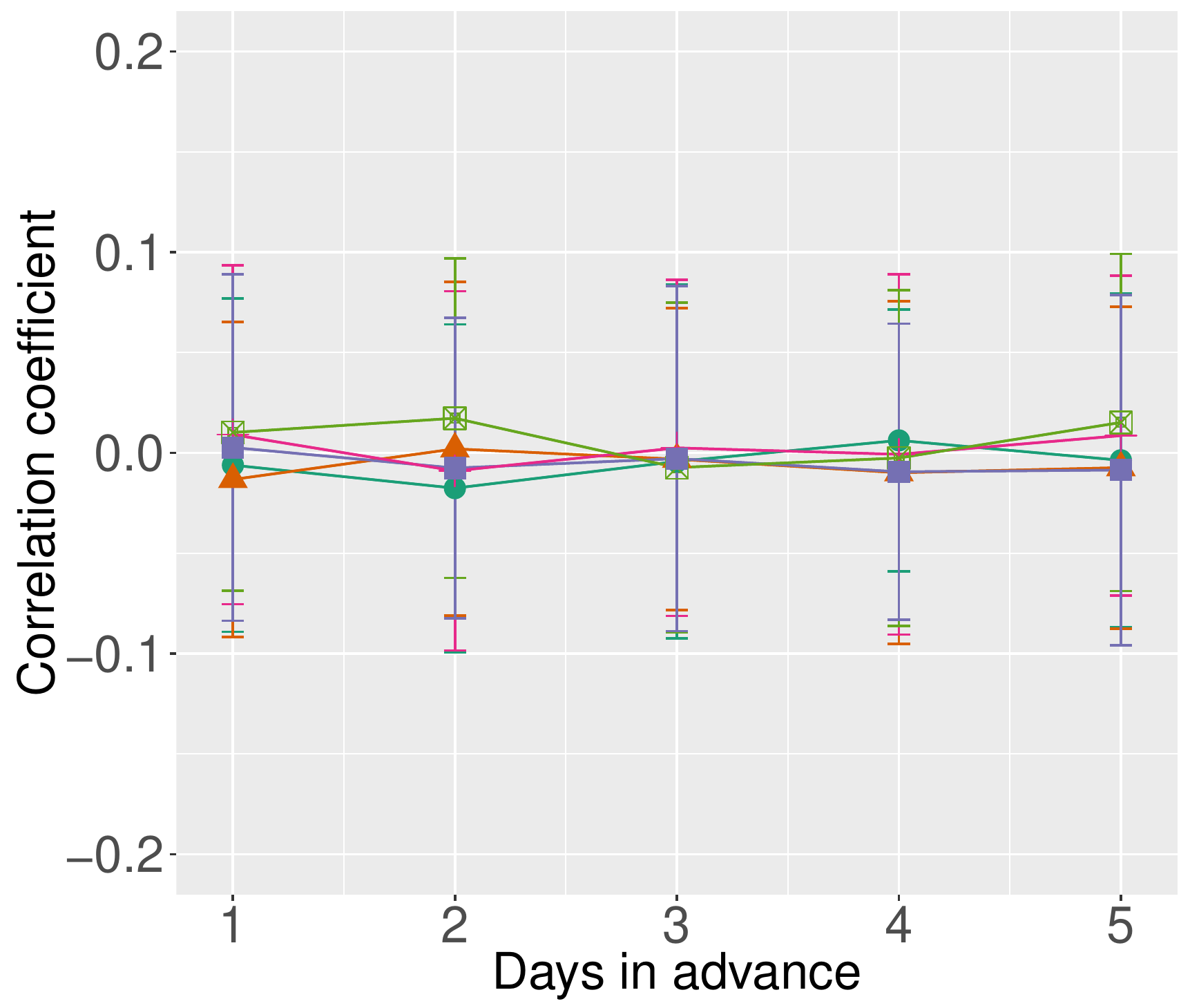}}
\hfil
\subfloat[$Y_{high}$ and $X$]{\includegraphics[width=0.28\textwidth]{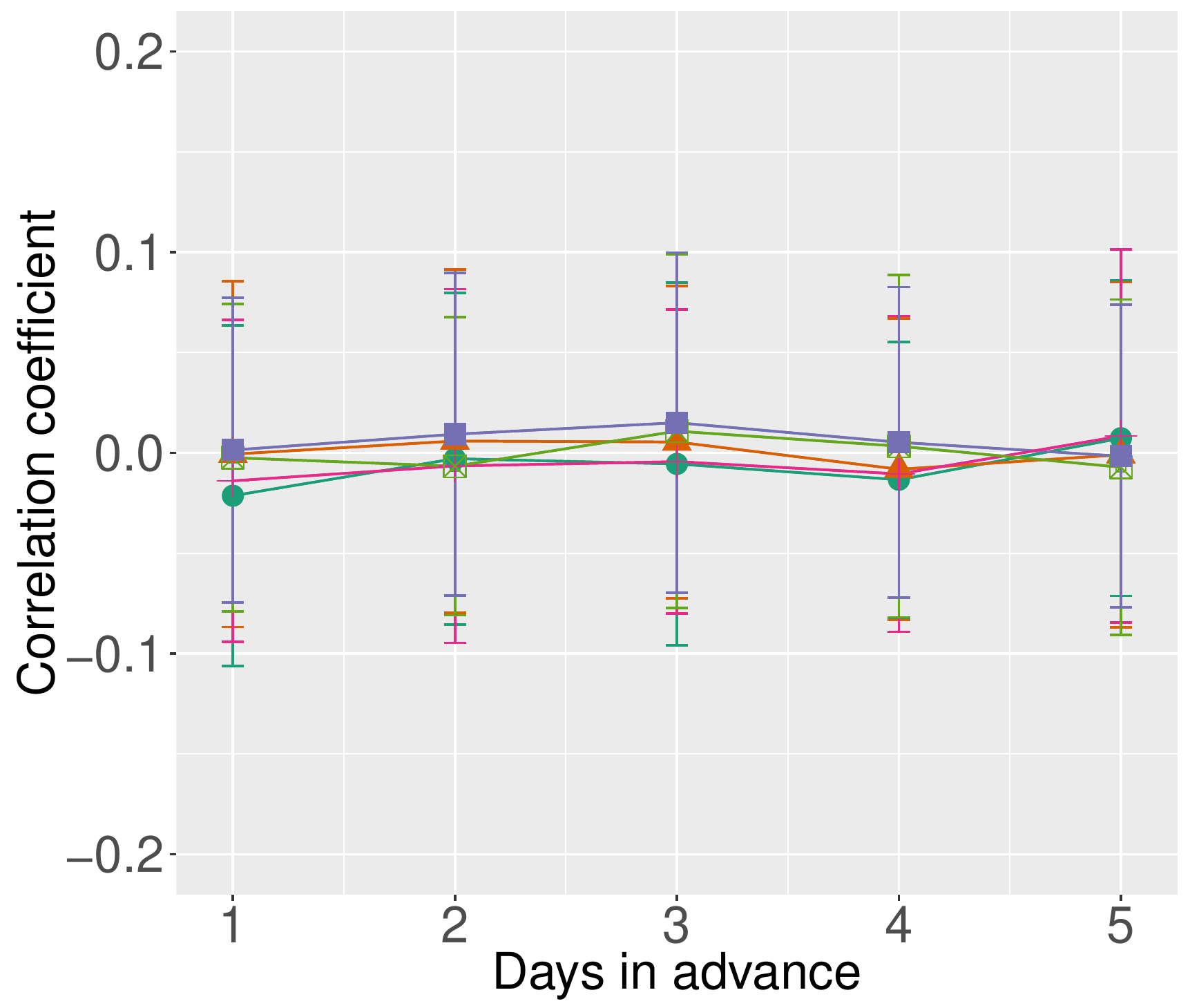}}
\hfil
\subfloat[$Y_{low}$ and $X$]{\includegraphics[width=0.28\textwidth]{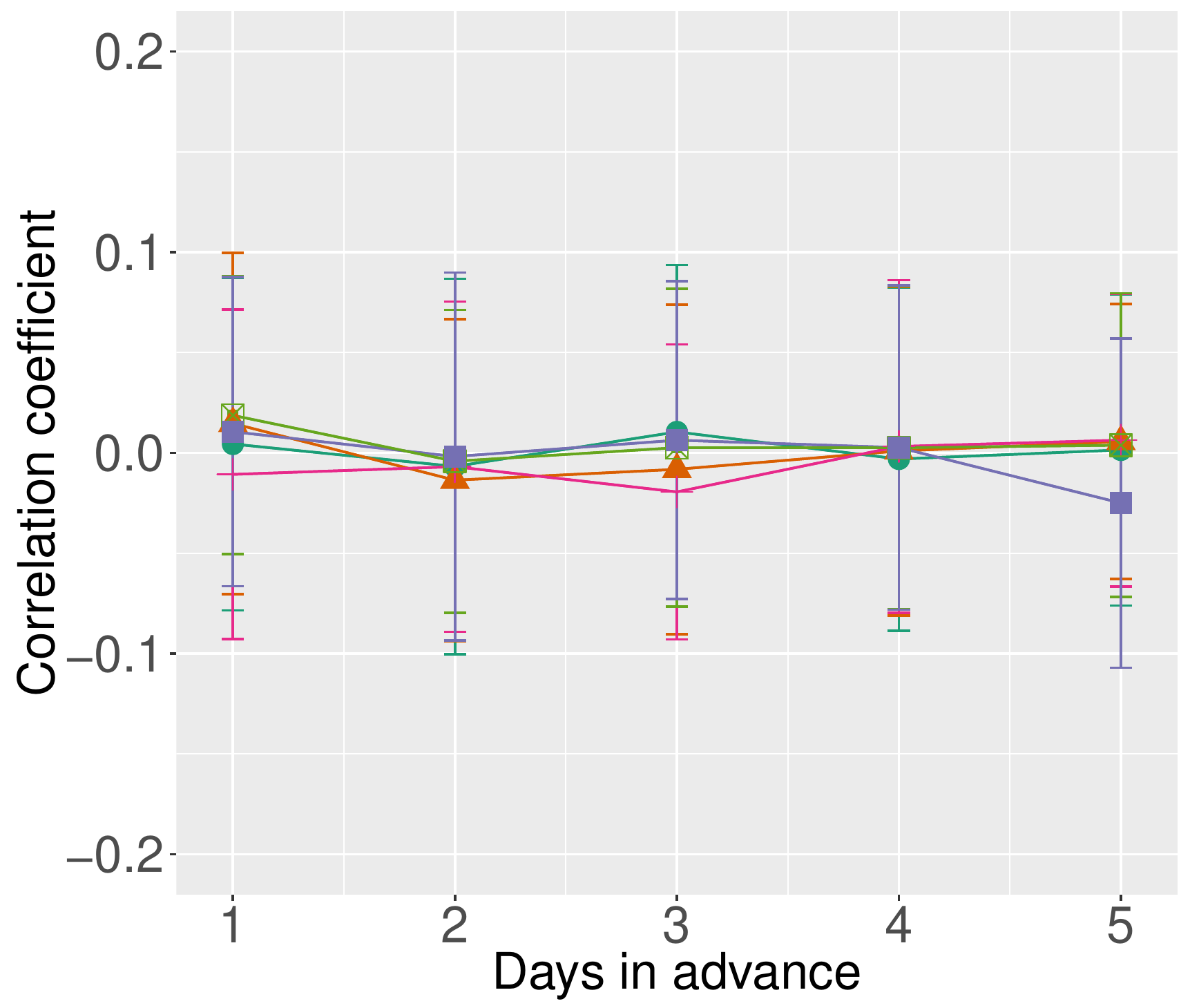}}
\hfil
\subfloat[$Y_{volume}$ and $X$]{\includegraphics[width=0.33\textwidth]{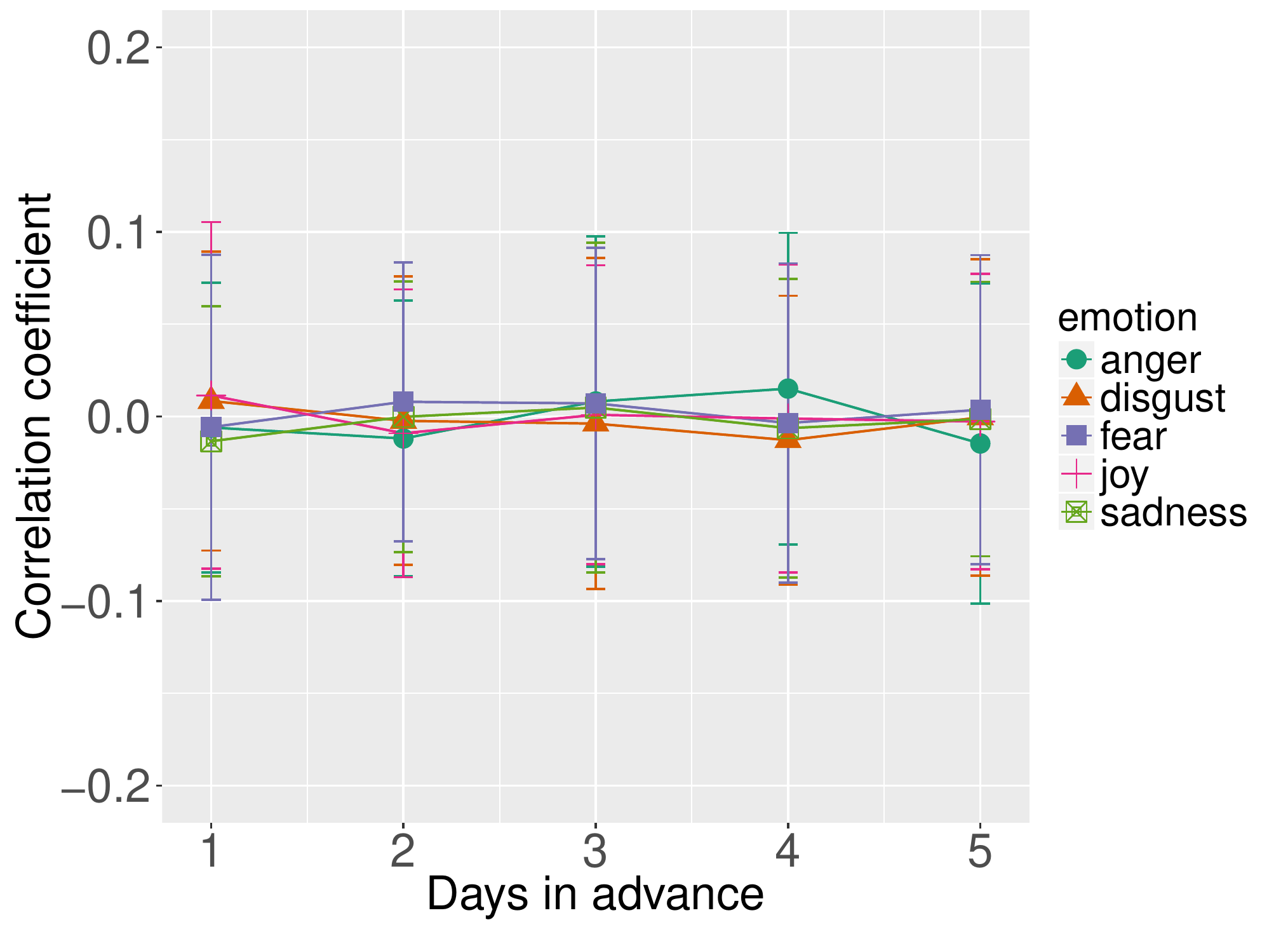}}
\caption{{\color{black}{Pearson correlation coefficient between five targets of stock market and shuffled online emotion data.}}}
\label{fig:shuffle}
\end{figure}

{\color{black}{For observing whether there are distinct differences of correlation coefficient between online emotions and the stock market, we use bootstrap resampling method to estimate the mean of correlation coefficients, as well as the error bars in the mean~\cite{hardle1991bootstrap,young1996jackknife}. Bootstrapping is often used with the purpose of deriving robust estimates of standard errors of a population parameter like correlation coefficients.}} Here, the emotion time series associated with stock market time series are sampled 100 times. In one time, we sample randomly 150 pairs (from 191 pairs in the training set) of data items respectively from emotion time series and stock market time series. We calculate 100 sampling results' correlation coefficients, and then obtain the mean values and standard deviations. Fig.~\ref{fig:error_bar} shows the means and error bars, which depicts sampling results' correlation coefficient. It can be seen that there are significant differences between different emotions. In addition, we randomly shuffle the time series 100 times and calculate the Pearson correlation coefficient {\color{black}{which is shown in Fig.~\ref{fig:shuffle}. As can be seen, the correlation coefficients become zero for all emotions, implying the significance of our previously disclosed correlations with the market attributes.}}

Inspecting the correlation coefficients above (shown in Fig.~\ref{fig:error_bar}), we set the threshold of correlation coefficient $\rho$ as 0.2 (the absolute value) and find some interesting and valuable results. The correlation between all online emotion time series (in $X$) and $Y_{close}$ is very low ($\rho<0.2$), which indicates little linear dependence between them. As to $Y_{open}$, the correlation coefficients with $X_{fear}$ (ahead of 1, 3, 4 and 5 days), $X_{joy}$ (ahead of 1 day) and $X_{disgust}$ (ahead of 1 and 5 days) are more than 0.2. $Y_{open}$ is negatively correlated with $X_{fear}$, positively correlated with $X_{joy}$ and $X_{disgust}$. As to $Y_{high}$, the correlation coefficients with $X_{joy}$ (ahead of 2 days) and $X_{sadness}$ (ahead of 2 days) are more than the threshold. $Y_{high}$ is negatively correlated with $X_{joy}$, positively correlated with $X_{sadness}$. $Y_{low}$ and 5 types of emotion time series have relatively high correlation, and the correlation coefficients between $Y_{low}$ and $X_{sadness}$ (ahead of 1 and 4 days) is the highest ($|\rho|>0.4$). $Y_{low}$ is negatively correlated with $X_{anger}$, $X_{disgust}$ and $X_{joy}$, positively correlated with $X_{sad}$ and $X_{fear}$. An interesting finding is that the correlation between $Y_{volume}$ and $X_{sadness}$ (no matter ahead of how many days) is unexpectedly high, correlation coefficients $\rho$ of which is more than 0.5. Besides, $Y_{volume}$ and other online emotion time series don't have a comparatively strong ($\rho>0.2$) correlation. 

{\color{black}{For further analysis, we inspect the correlation between targets and online emotions in past 30 trading days. It turns out that correlation decreases as time ahead increases. Besides, it is found that the predictive ability of the financial time series approaches the maximum as the five trading days in advance are considered~\cite{krauss2017deep}. Therefore, we take the online emotions of the past five days into the account in building the prediction models.}}

\subsection{Granger causality test of emotions and stock market}
\label{sec:granger}
Despite the correlation analysis, we also preform the causality test further on the training data. Here we apply the econometric approach named Granger causality test to study the relation between online emotions and the stock market. The Granger Causality Test is a statistical hypothesis test for determining whether one time series is functioning in forecasting another. One time series $x$ is said to Granger-cause $y$ if it can be shown that $x$ provides statistically significant information about future values of $y$, usually through a series of $t$-tests and $F$-tests on lagged values of $x$. We perform the analysis according to models shown in Eq~\ref{eq:granger_1} and~\ref{eq:granger_2} for the period from December 1st 2014 to September 16th 2015.
\begin{equation}
 y_{t} = \alpha + \sum_{i=1}^{n}\beta_{i}y_{t-i} + \epsilon_{t}, \\
\label{eq:granger_1}
\end{equation}

\begin{equation}
 y_{t} = \alpha + \sum_{i=1}^{n} \beta_{i}y_{t-i} + \sum_{i=1}^{n} \gamma_{i}x_{t-i} + \epsilon_{t}. \\ 
\label{eq:granger_2}
\end{equation}

The Granger causality test could select only two time series as inputs. We apply Granger causality test respectively on two groups: online emotion time series and stock market time series. Delaying time is set to 1, 2, 3, 4 and 5 days. According to different delaying time, we calculate the $p$-value to determine the results of hypothesis test. Here, the significance level is set to 5\%.

\begin{table*}[!t]
\centering
\caption{{\color{black}{Results of Granger causality test of online emotion and stock market time series. Only significant results are listed because of the limited space. $p$-value$<0.05$: *, $p$-value$<0.01$: **, $p$-value$<0.001$: ***.}}}
\label{tab:granger_causality_test}
\begin{tabular}{c|c|lllll}
\hline
\textbf{emotion}          & \textbf{lag (days)} & \textbf{Close} & \textbf{Open} & \textbf{High} & \textbf{Low} & \textbf{Volume} \\ \hline

\multirow{5}{*}{anger}    & 1             &                &               &               &              &                 \\
                          & 2             &                &               &               &              &                 \\
                          & 3             &                &               &               &              &                 \\
                          & 4             &                &               &               &              &                 \\
                          & 5             &                &               &               &              &                 \\ \hline
\multirow{5}{*}{disgust}  & 1 & $0.006^{**}$ & & & $0.032^{*}$ & \\ 
                          & 2 & $0.006^{**}$ & & & & \\
                          & 3 & & $0.007^{**}$ & & & \\
                          & 4 & & $0.019^{*}$ & & & \\
                          & 5 & & & $0.028^{*}$ & & \\ \hline
\multirow{5}{*}{joy}      & 1 & & $<0.001^{***}$ & $0.023^{*}$ & & \\ 
                          & 2 & & $<0.001^{***}$ & $0.030^{*}$ & & \\
                          & 3 & & $<0.001^{***}$ & $0.009^{**}$ & $0.006^{**}$ & \\
                          & 4 & & $<0.001^{***}$ & $0.039^{*}$ & $0.035^{*}$& \\
                          & 5 & & $<0.001^{***}$ & & & \\ \hline
\multirow{5}{*}{sadness}  & 1 & & & $0.012^{*}$ & $0.027^{*}$ & \\
                          & 2 & & & $0.022^{*}$ & & \\
                          & 3 & & & $0.030^{*}$ & & \\
                          & 4 & & & & & \\
                          & 5 & & & & & \\ \hline
\multirow{5}{*}{fear}     & 1 & & $<0.001^{***}$ & & & \\ 
                          & 2 & & $<0.001^{***}$ & & & \\
                          & 3 & & $<0.001^{***}$ & & & \\
                          & 4 & & $<0.001^{***}$ & & & \\
                          & 5 & & $<0.001^{***}$ & & & \\
\hline
\end{tabular}
\end{table*}

We list testing results whose $p$-value is {\color{black}{required to differ}} significant levels in Table~\ref{tab:granger_causality_test}. According to the results of Granger causality test, the null hypothesis, $X_{disgust} (lag=1,2)$ series do not predict $Y_{close}$, with a high level of confidence ($p$-value $<0.01$) can be rejected. However, the other emotions do not have causal relations with $Y_{close}$. $Y_{open}$ and $X_{joy}$ ($p$-value $<0.001$), $X_{fear}$ ($p$-value $<0.001$) and $X_{disgust}$  ($p$-value $<0.05$ or even $0.01$) have causal relations. $X_{joy}$, $X_{sadness}$ and $X_{disgust}$ have causal relations with $Y_{high}$ and $Y_{low}$ ($p$-value $<0.05$ or even $0.01$). At last, the results also suggest trading volume in stock market time series do not have significant causal relation with any emotion time series ($p$-value $\geq0.05$). It's surprising to find that $X_{anger}$ in online emotion time series does not have causal relation with any attribute of stock market in China.

The above analysis shows that $X_{disgust}$, $X_{joy}$, $X_{sadness}$ and $X_{fear}$ can be promising features for the stock prediction models, except for $X_{anger}$. {\color{black}{Regarding to users of different categories like F-level I and II, the Granger causality test demonstrates the similar relations between their emotions and the market. However, for the less emotional users in F-level III, it is surprisingly found that their anger shows a slightly significant Granger causality with the market. Nevertheless, because the trivial proportion of these users in investors, their anger accordingly contributes little to the prediction model based on aggregated emotions discussed later and hence we do not offer more discussions here.}}

\section{Predict the stock market}
\label{sec:predict}

Firstly, in this section, based on discretization methods, regression problems of predicting the stock market are converted to corresponding classification problems. Next, we perform linear and non-linear methods to solve the classification problems of stock market prediction. Eventually, the classification models are validated by 5-fold cross-validation on training set and we obtain a group of high-performance prediction models named SVM-ES.

For the prediction issue, we make use of the online emotion time series set (composed by shifted time series with different lags ranging from 1 to 5 for five emotions) or its subsets within the period from December 1st 2014 to September 9th 2015. Setting the longest lag to 5 trading days, the actual stock market time series are $Y$ from December 8th 2014 to September 16th 2015.

\subsection{Discretization of stock market data}
As illustrated in the previous sections, $Y_{close}, Y_{open}, Y_{high}$, $Y_{low}$ and $Y_{volume}$ are our targets of prediction in the stock market. Investors always just care for whether $Y_{close,i}$ (the element on $i$-th day in $Y_{close}$) are positive or negative, which will help investors make decisions to conduct stock transactions, and the binary classification (positive or negative) of $Y_{close}$ and $Y_{open}$ are also the part of our targets for prediction.

Besides, we convert regression problems of predicting five attributes in the stock market to classification problems by discretization methods through which we classify each of attributes to three categories. Specifically, $Y_{close}, Y_{open}, Y_{high}$, and $Y_{low}$ are divided into three categories: \textbf{bearish}(-1), \textbf{stable}(0) and \textbf{bullish}(1) represented by CLOSE, OPEN, HIGH and LOW below. $Y_{volume}$ are divide into three categories: \textbf{low}(-1), \textbf{normal}(0) and \textbf{high}(1) represented by VOLUME below.

\begin{figure*}[!t]
\centering
\subfloat[]{\includegraphics[width=0.32\textwidth]{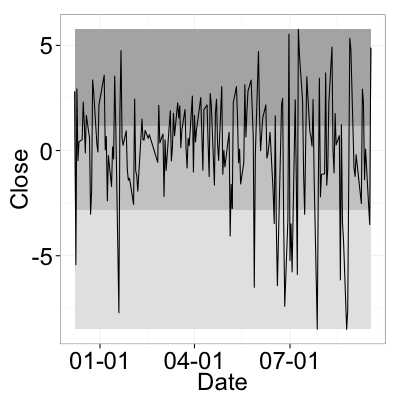}
\label{}}
\hfil
\subfloat[]{\includegraphics[width=0.32\textwidth]{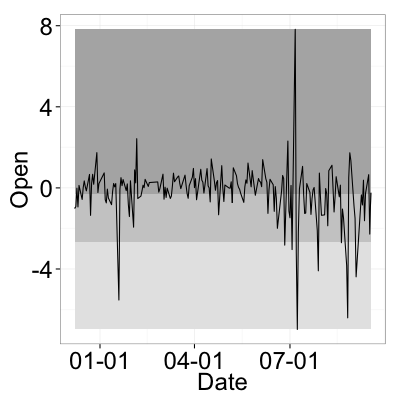}
\label{}}
\hfil
\subfloat[]{\includegraphics[width=0.32\textwidth]{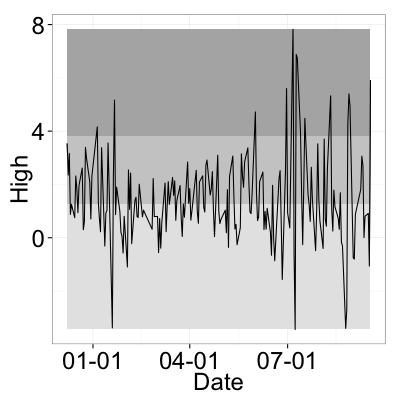}
\label{}}
\hfil
\subfloat[]{\includegraphics[width=0.32\textwidth]{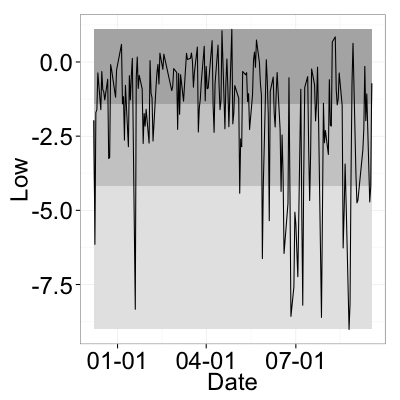}
\label{}}
\hfil
\subfloat[]{\includegraphics[width=0.35\textwidth]{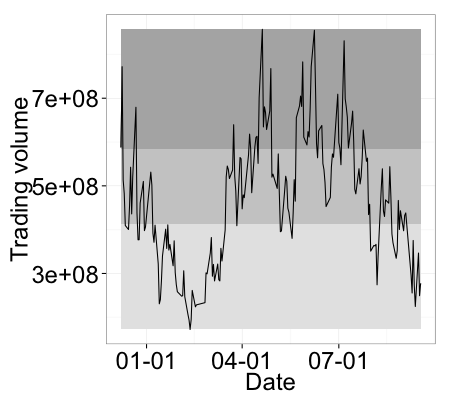}
\label{}}
\caption{Stock market time series and discretization results (by $K$-means) of $Y_{close}, Y_{open}, Y_{high}$, $Y_{low}$ and $Y_{volume}$.}
\label{fig:discretiztion}
\end{figure*}

The discretization of five attributes in the stock market is conducted by two methods: equal frequency and $K$-means clustering. Equal frequency discretization is a simple but effective method that we sort items from large to small then cut them into 3 clusters of even size. $K$-means clustering, another method we use, is popular for cluster analysis in data mining. In this paper, $K$-means clustering aims to partition observations of the stock market into 3 clusters in which each observation belongs to the cluster with the nearest distance. The results of three categories discretization by $K$-means are shown in Fig.~\ref{fig:discretiztion} with 3 different grey levels.

Intuitively, as compared to the approach of equal frequency, the discretization based on $K$-means is more flexible and adjustable to the dynamic of the market. The categories which it generates can better reflect the actual market status {\color{black}{(it is hard to say that the market will rise, drop or keep stable with the same probability)}} and thus can offer us a better benchmark to test the prediction results. 

\subsection{Classification model for stock prediction}
In this paper, we perform machine learning methods, Logistic Regression (linear) and Support Vector Machine (non-linear), to solve the classification problems for stock prediction. These methods are both popular for training binary or multiple classification. To predict the categories $(-1, 0, 1)$ or $(0, 1)$ (just for CLOSE and OPEN) of $Y$ on $i$-th day, the input attributes of our Logistic Regression model (LR) and Support Vector Machine model (SVM) include only online emotion values of the past 5 days or a subset of them, except for other variables in the field of finance. We adapt 5-fold cross-validation to examine the accuracies of models.

At the outset, we consider all five emotions of the past 5 days as the input attributes of LR and SVM. The accuracies of models by 5-fold cross-validation are shown in Table~\ref{tab:cv_result} (3-categories and 2-categories). For the classification problem in this paper, the performance of SVM is always better than that of LR. Therefore, we conjecture that, relation between online emotions and the stock market is not simply linear, and non-linear prediction models are suited to model the complex interactions of factors involved in shaping financial market values.

While 3-categories discretization $(-1, 0, 1)$ results in the stock market as predicted targets, $K$-means clustering is always better than equal frequency discretization. In other words, the accuracies of models by 5-fold cross-validation, of which predicted targets are the results by $K$-means clustering, are relatively higher. Considering the categories generated by $K$-means discretization better represent the market status, we can conclude that our models indeed capture the essence of the stock fluctuation. {\color{black}{Note that for other $K$ values like 5 or 7, it is hard to establish competent prediction models because of the sparsity and hence the results are not reported. In fact, in realistic applications, it is sufficient for investors to derive decisions from the prediction model of $K=3$.}}

\begin{table}
\setlength{\abovecaptionskip}{2pt}
\caption{Accuracy of 5-fold cross-validation for 3-categories and 2-categories prediction models.}
\centering
\begin{tabular}{ccccccc}
\hline
\multirow{2}*{Target (3)} & 
\multicolumn{2}{c}{Equal frequency} & 
\multicolumn{3}{c}{$K$-means} \\
\cmidrule(lr){2-3}\cmidrule(lr){4-6}
& LR & SVM & LR & \textbf{SVM} & \textbf{SVM-ES}\\
\hline
CLOSE & 34.0\%  & 43.5\%  & 52.9\%  & \textbf{58.1}\%  & 57.6\% \\
OPEN  & 37.7\%  & 44.0\%  & 53.4\%  & 61.3\%  & \textbf{64.4}\% \\
HIGH  & 36.7\%  & 39.3\%  & 48.7\%  & 53.4\%  & \textbf{54.5}\% \\
LOW   & 42.4\%  & 49.2\%  & 57.0\%  & 63.4\%  & \textbf{64.4}\% \\
VOLUME& 50.8\%  & 63.9\%  & 53.4\%  & \textbf{67.0}\%  & 66.5\% \\
\hline
\label{tab:cv_result}
\end{tabular}

\begin{tabular}{ccccc}
\hline
Target (2) & LR & SVM & SVM-ES \\
\hline
CLOSE & 58.1\% & \textbf{61.3}\% & 60.2\% \\
OPEN  & 58.1\% & \textbf{66.0}\% & 64.9\% \\
\hline
\label{tab:two_classes_result}
\end{tabular}
\end{table}

However, recalling the correlation analysis and Granger causality test of online emotions and the stock market, not all the emotions play roles on predicting the stock market and the analysis results should be used for the feature selection. Consequently, we build support vector machine model based emotions selected (SVM-ES) for stock prediction (discretized by $K$-means). The input attributes are based on analysis results of Granger causality test and Pearson correlation. We select $X_{disgust}$ (ahead of 1, 2 days) for the SVM-ES to predict CLOSE, $X_{fear}$ (ahead of 1-5 days), $X_{joy}$ (ahead of 1-5 day) and $X_{disgust}$ (ahead of 3 and 4 days) as the input attributes for predicting OPEN, $X_{joy}$ (ahead of 1-4 days), $X_{sadness}$ (ahead of 1-3 day) and $X_{disgust}$ (ahead of 5 days) as the input attributes for predicting HIGH, and $X_{sadness}$ (ahead of 1 day), $X_{joy}$ (ahead of 1-3 day) and $X_{disgust}$ (ahead of 5 days) as the input attributes for predicting LOW. Correlation analysis of $Y_{volume}$ indicates that $Y_{volume}$ and $X_{sadness}$ (ahead of 1-5 days) have the strongest correlation ($\rho>0.5$) among all online emotions, however, just using sadness as the learning feature surprisingly cannot guarantee the expected performance. Thus, we try to select $X_{sadness}$ (ahead 1-5 days) and $X_{fear}$ (ahead 1-5 days) which is the second strongest relation with $Y_{volume}$ as the input attributes to predict VOLUME.

After adjusting and fixing the input attributes of SVM-ES, we train the models for stock prediction. The last column of Table~\ref{tab:cv_result} shows the accuracy of 5-fold cross-validation, respectively for 3-categories and 2-categories classification models. There are slight differences in performance between SVM-ES and the SVM trained using all the emotions, indicating emotions selected are playing dominant roles in forecasting the market. It is noteworthy that input attributes of all the SVM-ES don't include anger and it's surprising that anger shown in online social media possesses the weakest correlation or even no relation with the Chinese stock market.

From Table~\ref{tab:cv_result} it should be also noted that, emotions selected can boost the classification results attributes like OPEN, HIGH and LOW, while for CLOSE and VOLUME, SVM with all emotions as features is still the most competent solution, with slight increment (around 1\%) to SVM-ES (few attributes of input). This result explains that emotions except for input attributes of SVM-ES have very weak effects on the stock market prediction.

\subsection{Evaluation in realistic application}
\begin{figure}
\centering
 \includegraphics[width=10cm]{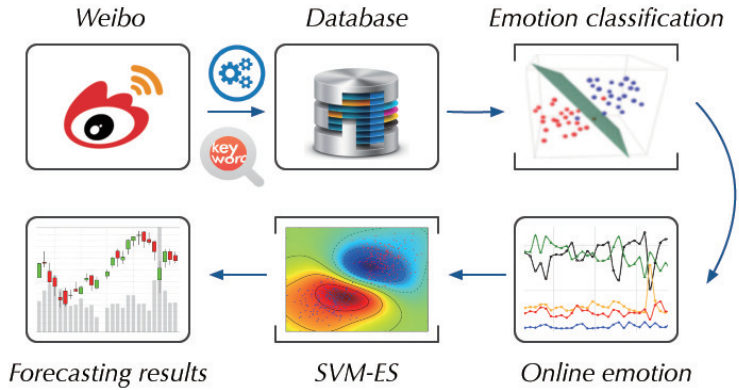}
\caption{Framework of realistic application for stock prediction based on SVM-ES.}
\label{fig:system}
\end{figure}

{\color{black}{To evaluate our prediction models}}, we sustain collecting stock-relevant tweets on Weibo from September 17th to December 7th in 2015. We process these tweets and obtain online emotion time series as our testing set. Then we apply our classification models SVM-ES for stock prediction in the realistic Chinese stock market and we can get the daily predictions of five attributes before the market open. Framework of realistic application based on SVM-ES is demonstrated in Fig.~\ref{fig:system}. We evaluate the stock market prediction application and the accuracy is shown in Table~\ref{tab:test_SVMES}. It turns out that the model achieves the high prediction performance, especially with accuracy of 64.15\% for the intra-day highest index (3-categories) and the accuracy of 60.38\% for the trading volume (3-categories).

{\color{black}{To further examine the robustness of predictive power of online emotions, we compare the prediction performance of the models which take emotions time series or market time series as the features. Inspired with traditional financial time series forecasting researches~\cite{Kara20115311,kim2003financial}, the baseline SVM models based on market return, named SVM-MR, are established. The input attributes of SVM-MR are index return in the past five days, which are represented by $Close_{i-lag}$ ($lag=1,2,3,4$ and $5$). Besides, the prediction targets and the periods of training set of two groups of models are identical. It can be seen from Table~\ref{tab:test_SVMES} that the performance of SVM-ES models is always better than baselines. We can conclude that online emotions offer more predictive power than simple financial time series.}}

{\color{black}{We also evaluate the predictive power of online emotions extracted from different categories of users. Using the identical method with SVM-ES, we build SVM models based on five emotions from different user types. We focus on two important targets, including closing index (3-categories and 2-categories) and trading volume (3-categories). The results in Table.~\ref{tab:SVM_F-level} show that the performance of models based on emotions from F-level I and II is better than from F-level III. With closing index as the target, the accuracy of the model based on F-level III is even worse than the random result. The phenomenon could be explained by ``Efficient Market Hypothesis''. According to hypothesis, the stock market will follow a random walk pattern and cannot be predicted while every investor in market is rational. Therefore, it is not available to predict the market only by using online emotions of rational investment experts. In comparison, the performance of SVM-ES, which considers emotions from various kinds of users, is excellent. It concludes that the aggregation of emotions from all investors in Weibo have more predictive power than any single investor category. An explanation for this phenomenon can be the market is affected by emotions and behaviors from categories of investors. Online emotions in whole social media, rather than its subset, can reflect the overall market status.}}

\begin{table}
\caption{{\color{black}{Accuracy of SVM-MR and SVM-ES on realistic application.}}}
\centering
\begin{tabular}{@{}ccc@{}}
\hline
Target    & SVM-MR  & \textbf{SVM-ES}  \\ \hline
CLOSE(3)  & 45.28\% & \textbf{56.60\%} \\
OPEN(3)   & 43.39\% & \textbf{43.40\%} \\
HIGH(3)   & 49.06\% & \textbf{64.15\%} \\
LOW(3)    & 47.17\% & \textbf{56.60\%} \\
VOLUME(3) & 56.60\% & \textbf{60.38\%} \\
CLOSE(2)  & 52.83\% & \textbf{60.38\%} \\
OPEN(2)   & 49.06\% & \textbf{56.60\%} \\
\hline
\label{tab:test_SVMES}
\end{tabular}
\end{table}

\begin{table}
\centering
\caption{{\color{black}{Accuracy of SVM models based on emotions from different categories of users.}}}
\label{tab:SVM_F-level}
\begin{tabular}{@{}cccc@{}}
\hline
Target    & \textbf{F-level I} & \textbf{F-level I} & F-level III \\ \hline
CLOSE(3)  & \textbf{54.72\%}   & 50.94\%   & 26.42\%     \\
VOLUME(3) & \textbf{62.26\%}   & 60.38\%            & 56.60\%     \\
CLOSE(2)  & 50.94\%            & \textbf{52.83\%}   & 47.17\%     \\ \hline
\end{tabular}
\end{table}

\section{Conclusion}
\label{sec:conclusion}

{\color{black}{Focusing on the stock market in China}}, we collect massive stock-relevant tweets in Weibo and label them with five categories of sentiments. By defining the emotional volatility, it is the first to reveal that inexperienced investors (over 98\%) are more sensible to the market fluctuations than the experienced or institutional ones, suggesting the Chinese market is emotional. Then the correlation analysis and Granger causality test are performed. {\color{black}{We suggest there may be significant potential for short-term prediction based on the online emotions of the past five days. Both analysis results show that disgust, joy, sadness and fear can be used to predict the price and volume of the stock mark.}} Based on these, we establish several models to predict the closing index, the opening index, the intra-day highest index, the intra-day lowest index and trading volume. The results show that our model SVM-ES can outperform baseline solutions. Finally, we also testify its performance in the realistic application and compare it with the model based on market return. In conclusion, {\color{black}{our findings confirm that the stock market in China can be predicted by various online emotions.}}

{\color{black}{Indeed, social media provides a valuable data source for researchers and companies where publicly available tweets can be retrieved as well as analyzed with the help of machine learning technologies. In this study, we argue that from the perspective of computational social science, tweets in Weibo can be excellent proxies in probing the financial market and corresponding emotions. In addition, to our best knowledge, we are the first to establish that, there are significant differences of emotional responses from types of online users while facing the changing market. Therefore, the behavioral differences should be considered when studying the relationship between social media and real world.}}

{\color{black}{According to the competent performance of models in realistic application, our results have strong implications for investors as well as the entire stock market in China. The financial industry might integrate online emotions into traditional prediction models to make better trading decisions. Even the combination of sentiment analysis with classical capital market models would be an interesting field and promising direction in behavioral financial.}}

This study has inevitable limitations, which might be interesting directions in the future work. For example, how the connection between emotion and market evolves with time is not fully discussed, however, which could help to design incremental learning schemes in realistic applications. Meanwhile, investors with stable emotion volatility can be selected to automatically recommend strategies because of their rich experience and expertise.


\bibliographystyle{splncs03}

\end{document}